\documentclass[aps,pra,twocolumn,amsmath,amssymb,nofootinbib,superscriptaddress]{revtex4-1}

\usepackage{times}
\usepackage[pdftex]{graphicx}
\usepackage{dcolumn}
\usepackage{bm}
\usepackage{amsmath}
\usepackage{indentfirst}
\usepackage{float}
\usepackage[colorlinks]{hyperref}
\usepackage[dvipsnames]{xcolor}
\usepackage{xcolor}

\usepackage{subfigure}
\usepackage{algorithm}
\usepackage{algorithmicx}
\usepackage{algpseudocode}
\usepackage{amsmath}
\usepackage{verbatim}

\begin{document}

\title{Strong quantum computational advantage using a superconducting quantum processor}
%Characterizing a 66-Qubit Quantum Processor via Random Quantum Circuit Benchmarking
%Enhanced Quantum Computational Advantage with a Superconducting Quantum Processor
\author{Yulin Wu}
\affiliation{Hefei National Laboratory for Physical Sciences at the Microscale and Department of Modern Physics, University of Science and Technology of China, Hefei 230026, China}
\affiliation{Shanghai Branch, CAS Center for Excellence in Quantum Information and Quantum Physics, University of Science and Technology of China, Shanghai 201315, China}
\affiliation{Shanghai Research Center for Quantum Sciences, Shanghai 201315, China}
\author{Wan-Su Bao}
\affiliation{Henan Key Laboratory of Quantum Information and Cryptography, Zhengzhou 450000, China}
\author{Sirui Cao}
\author{Fusheng Chen}
\author{Ming-Cheng Chen}
\affiliation{Hefei National Laboratory for Physical Sciences at the Microscale and Department of Modern Physics, University of Science and Technology of China, Hefei 230026, China}
\affiliation{Shanghai Branch, CAS Center for Excellence in Quantum Information and Quantum Physics, University of Science and Technology of China, Shanghai 201315, China}
\affiliation{Shanghai Research Center for Quantum Sciences, Shanghai 201315, China}
\author{Xiawei Chen}
\affiliation{Shanghai Branch, CAS Center for Excellence in Quantum Information and Quantum Physics, University of Science and Technology of China, Shanghai 201315, China}
\author{Tung-Hsun Chung}
\author{Hui Deng}
\affiliation{Hefei National Laboratory for Physical Sciences at the Microscale and Department of Modern Physics, University of Science and Technology of China, Hefei 230026, China}
\affiliation{Shanghai Branch, CAS Center for Excellence in Quantum Information and Quantum Physics, University of Science and Technology of China, Shanghai 201315, China}
\affiliation{Shanghai Research Center for Quantum Sciences, Shanghai 201315, China}
\author{Yajie Du}
\affiliation{Shanghai Branch, CAS Center for Excellence in Quantum Information and Quantum Physics, University of Science and Technology of China, Shanghai 201315, China}
\author{Daojin Fan}
\author{Ming Gong}
\author{Cheng Guo}
\author{Chu Guo}
\author{Shaojun Guo}
\author{Lianchen Han}
\affiliation{Hefei National Laboratory for Physical Sciences at the Microscale and Department of Modern Physics, University of Science and Technology of China, Hefei 230026, China}
\affiliation{Shanghai Branch, CAS Center for Excellence in Quantum Information and Quantum Physics, University of Science and Technology of China, Shanghai 201315, China}
\affiliation{Shanghai Research Center for Quantum Sciences, Shanghai 201315, China}
\author{Linyin Hong}
\affiliation{QuantumCTek Co., Ltd., Hefei 230026, China}
\author{He-Liang Huang}
\affiliation{Hefei National Laboratory for Physical Sciences at the Microscale and Department of Modern Physics, University of Science and Technology of China, Hefei 230026, China}
\affiliation{Shanghai Branch, CAS Center for Excellence in Quantum Information and Quantum Physics, University of Science and Technology of China, Shanghai 201315, China}
\affiliation{Shanghai Research Center for Quantum Sciences, Shanghai 201315, China}
\affiliation{Henan Key Laboratory of Quantum Information and Cryptography, Zhengzhou 450000, China}
\author{Yong-Heng Huo}
\affiliation{Hefei National Laboratory for Physical Sciences at the Microscale and Department of Modern Physics, University of Science and Technology of China, Hefei 230026, China}
\affiliation{Shanghai Branch, CAS Center for Excellence in Quantum Information and Quantum Physics, University of Science and Technology of China, Shanghai 201315, China}
\affiliation{Shanghai Research Center for Quantum Sciences, Shanghai 201315, China}
\author{Liping Li}
\affiliation{Shanghai Branch, CAS Center for Excellence in Quantum Information and Quantum Physics, University of Science and Technology of China, Shanghai 201315, China}
\author{Na Li}
\author{Shaowei Li}
\author{Yuan Li}
\author{Futian Liang}
\affiliation{Hefei National Laboratory for Physical Sciences at the Microscale and Department of Modern Physics, University of Science and Technology of China, Hefei 230026, China}
\affiliation{Shanghai Branch, CAS Center for Excellence in Quantum Information and Quantum Physics, University of Science and Technology of China, Shanghai 201315, China}
\affiliation{Shanghai Research Center for Quantum Sciences, Shanghai 201315, China}
\author{Chun Lin}
\affiliation{Shanghai Institute of Technical Physics, Chinese Academy of Sciences, Shanghai 200083, China}
\author{Jin Lin}
\author{Haoran Qian}
\affiliation{Hefei National Laboratory for Physical Sciences at the Microscale and Department of Modern Physics, University of Science and Technology of China, Hefei 230026, China}
\affiliation{Shanghai Branch, CAS Center for Excellence in Quantum Information and Quantum Physics, University of Science and Technology of China, Shanghai 201315, China}
\affiliation{Shanghai Research Center for Quantum Sciences, Shanghai 201315, China}
\author{Dan Qiao}
\affiliation{Shanghai Branch, CAS Center for Excellence in Quantum Information and Quantum Physics, University of Science and Technology of China, Shanghai 201315, China}
\author{Hao Rong}
\author{Hong Su}
\author{Lihua Sun}
\affiliation{Hefei National Laboratory for Physical Sciences at the Microscale and Department of Modern Physics, University of Science and Technology of China, Hefei 230026, China}
\affiliation{Shanghai Branch, CAS Center for Excellence in Quantum Information and Quantum Physics, University of Science and Technology of China, Shanghai 201315, China}
\affiliation{Shanghai Research Center for Quantum Sciences, Shanghai 201315, China}
\author{Liangyuan Wang}
\affiliation{Shanghai Branch, CAS Center for Excellence in Quantum Information and Quantum Physics, University of Science and Technology of China, Shanghai 201315, China}
\author{Shiyu Wang}
\author{Dachao Wu}
\author{Yu Xu}
\affiliation{Hefei National Laboratory for Physical Sciences at the Microscale and Department of Modern Physics, University of Science and Technology of China, Hefei 230026, China}
\affiliation{Shanghai Branch, CAS Center for Excellence in Quantum Information and Quantum Physics, University of Science and Technology of China, Shanghai 201315, China}
\affiliation{Shanghai Research Center for Quantum Sciences, Shanghai 201315, China}
\author{Kai Yan}
\affiliation{Shanghai Branch, CAS Center for Excellence in Quantum Information and Quantum Physics, University of Science and Technology of China, Shanghai 201315, China}
\author{Weifeng Yang}
\affiliation{QuantumCTek Co., Ltd., Hefei 230026, China}
\author{Yang Yang}
\affiliation{Shanghai Branch, CAS Center for Excellence in Quantum Information and Quantum Physics, University of Science and Technology of China, Shanghai 201315, China}
\author{Yangsen Ye}
\affiliation{Hefei National Laboratory for Physical Sciences at the Microscale and Department of Modern Physics, University of Science and Technology of China, Hefei 230026, China}
\affiliation{Shanghai Branch, CAS Center for Excellence in Quantum Information and Quantum Physics, University of Science and Technology of China, Shanghai 201315, China}
\affiliation{Shanghai Research Center for Quantum Sciences, Shanghai 201315, China}
\author{Jianghan Yin}
\affiliation{Shanghai Branch, CAS Center for Excellence in Quantum Information and Quantum Physics, University of Science and Technology of China, Shanghai 201315, China}
\author{Chong Ying}
\author{Jiale Yu}
\author{Chen Zha}
\author{Cha Zhang}
\affiliation{Hefei National Laboratory for Physical Sciences at the Microscale and Department of Modern Physics, University of Science and Technology of China, Hefei 230026, China}
\affiliation{Shanghai Branch, CAS Center for Excellence in Quantum Information and Quantum Physics, University of Science and Technology of China, Shanghai 201315, China}
\affiliation{Shanghai Research Center for Quantum Sciences, Shanghai 201315, China}
\author{Haibin Zhang}
\affiliation{Shanghai Branch, CAS Center for Excellence in Quantum Information and Quantum Physics, University of Science and Technology of China, Shanghai 201315, China}
\author{Kaili Zhang}
\author{Yiming Zhang}
\affiliation{Hefei National Laboratory for Physical Sciences at the Microscale and Department of Modern Physics, University of Science and Technology of China, Hefei 230026, China}
\affiliation{Shanghai Branch, CAS Center for Excellence in Quantum Information and Quantum Physics, University of Science and Technology of China, Shanghai 201315, China}
\affiliation{Shanghai Research Center for Quantum Sciences, Shanghai 201315, China}
\author{Han Zhao}
\affiliation{Shanghai Branch, CAS Center for Excellence in Quantum Information and Quantum Physics, University of Science and Technology of China, Shanghai 201315, China}
\author{Youwei Zhao}
\affiliation{Hefei National Laboratory for Physical Sciences at the Microscale and Department of Modern Physics, University of Science and Technology of China, Hefei 230026, China}
\affiliation{Shanghai Branch, CAS Center for Excellence in Quantum Information and Quantum Physics, University of Science and Technology of China, Shanghai 201315, China}
\affiliation{Shanghai Research Center for Quantum Sciences, Shanghai 201315, China}
\author{Liang Zhou}
\affiliation{QuantumCTek Co., Ltd., Hefei 230026, China}
\author{Qingling Zhu}
\author{Chao-Yang Lu}
\author{Cheng-Zhi Peng}
\author{Xiaobo Zhu}
\author{Jian-Wei Pan}

\affiliation{Hefei National Laboratory for Physical Sciences at the Microscale and Department of Modern Physics, University of Science and Technology of China, Hefei 230026, China}
\affiliation{Shanghai Branch, CAS Center for Excellence in Quantum Information and Quantum Physics, University of Science and Technology of China, Shanghai 201315, China}
\affiliation{Shanghai Research Center for Quantum Sciences, Shanghai 201315, China}

\date{\today}


\begin{abstract}
Scaling up to a large number of qubits with high-precision control is essential in the demonstrations of quantum computational advantage to exponentially outpace the classical hardware and algorithmic improvements. Here, we develop a two-dimensional programmable superconducting quantum processor, \textit{Zuchongzhi}, which is composed of 66 functional qubits in a tunable coupling architecture. To characterize the performance of the whole system, we perform random quantum circuits sampling for benchmarking, up to a system size of 56 qubits and 20 cycles. The computational cost of the classical simulation of this task is estimated to be 2-3 orders of magnitude higher than the previous work on 53-qubit Sycamore processor [Nature \textbf{574}, 505 (2019)]. We estimate that the sampling task finished by \textit{Zuchongzhi} in about 1.2 hours will take the most powerful supercomputer at least 8 years. Our work establishes an unambiguous quantum computational advantage that is infeasible for classical computation in a reasonable amount of time. The high-precision and programmable quantum computing platform opens a new door to explore novel many-body phenomena and implement complex quantum algorithms.
\end{abstract}

\maketitle

\section{Introduction}
In the past years, encouraging progress has been made in the physical realizations of quantum computers~\cite{bruzewicz2019trapped, huang2020superconducting, slussarenko2019photonic,gong2021quantum}, indicating a transition of quantum computing from a theoretical picture to a nascent technology. A major milestone along the way is the demonstration of quantum computational advantage, which is also known as quantum supremacy. It is defined by a quantum device that can implement a well-defined task overwhelmingly faster than any classical computer to an extent that no classical computer can complete the task within a reasonable amount of time.

To this end, recent experiments using 53 superconducting qubits and 76 photons have provided strong evidence to demonstrate the quantum computational advantage and subsequently disprove the extended Church-Turing thesis~\cite{aaronson2011computational,boixo2018characterizing,bouland2019complexity,aaronson2016complexity}. Due to continuous improvements in the classical algorithm and hardware ~\cite{
pednault2019leveraging,HuangChen2020,pan2021simulating} to compete the quantum computers, the demonstration of quantum computational advantage is not a single-shot achievement but the quantum hardware has to be upgraded. It should be emphasized that the increase of qubits is expected to exponentially outpace the classical performance.

Simultaneously increasing the number of qubits and high-fidelity quantum logic gates are also crucial for the rapid development of noisy intermediate scale quantum (NISQ) technology~\cite{preskill2018quantum} and the demonstration of logic qubit through surface code error correction
~\cite{fowler2012surface,shor1996fault,erhard2021entangling, andersen2020repeated, marques2021logical,chen2021exponential}. Indeed, a wide range of near-term applications are being investigated, including quantum chemistry~\cite{google2020hartree,mcardle2020quantum,aspuru2005simulated}, quantum many-body physics~\cite{bernien2017probing,zhang2017observation, zhu2021observation,chen2021observation,zha2020ergodic}, and quantum machine learning~\cite{huang2020experimental,havlivcek2019supervised,harrigan2021quantum,huang2018demonstration, saggio2021experimental,cong2019quantum}.

Scaling up high-fidelity superconducting quantum processors faces major challenges in the chip fabrication and qubit control. In this work, we make progress toward building a larger-scale and high-performance superconducting quantum computing system, named \textit{Zuchongzhi}. The quantum processor is designed and fabricated with a two-dimensional and tunable coupling architecture, which contains a total of 66 qubits. High-fidelity single-qubit gate (average 99.86\%) and two-qubit gate (average 99.41\%), as well as readout (average 95.48\%), are achieved with this processor, while performing simultaneous gate operations on multiple qubits. We use random quantum circuit sampling~\cite{boixo2018characterizing} as a metric to evaluate the overall power of the quantum processor. Experimental results show that our processor is able to complete the sampling task with a system size up to 56 qubits and 20 cycles. We estimate that the classical computational overhead to simulate \textit{Zuchongzhi} is 2-3 orders of magnitude higher than the task implemented on Google's 53-qubit Sycamore processor~\cite{arute2019quantum}. Therefore, our experiment unambiguously established a computational task that can be completed by a quantum computer in 1.2 hours but will take at least an unreasonable time for any supercomputers.

%Furthermore, our high-fidelity quantum processor has a scalable architecture that is compatible with surface code error correction, which can act as the test bed for fault tolerant quantum computing, and enables further development of valuable near-term applications. The achieved results suggest that our powerful quantum computing prototype would be an excellent candidate system for exploring a variety of various near-term applications.

\section{High-Performance Quantum Processor}

\begin{figure}[tbp]
\begin{center}
\includegraphics[width=\linewidth]{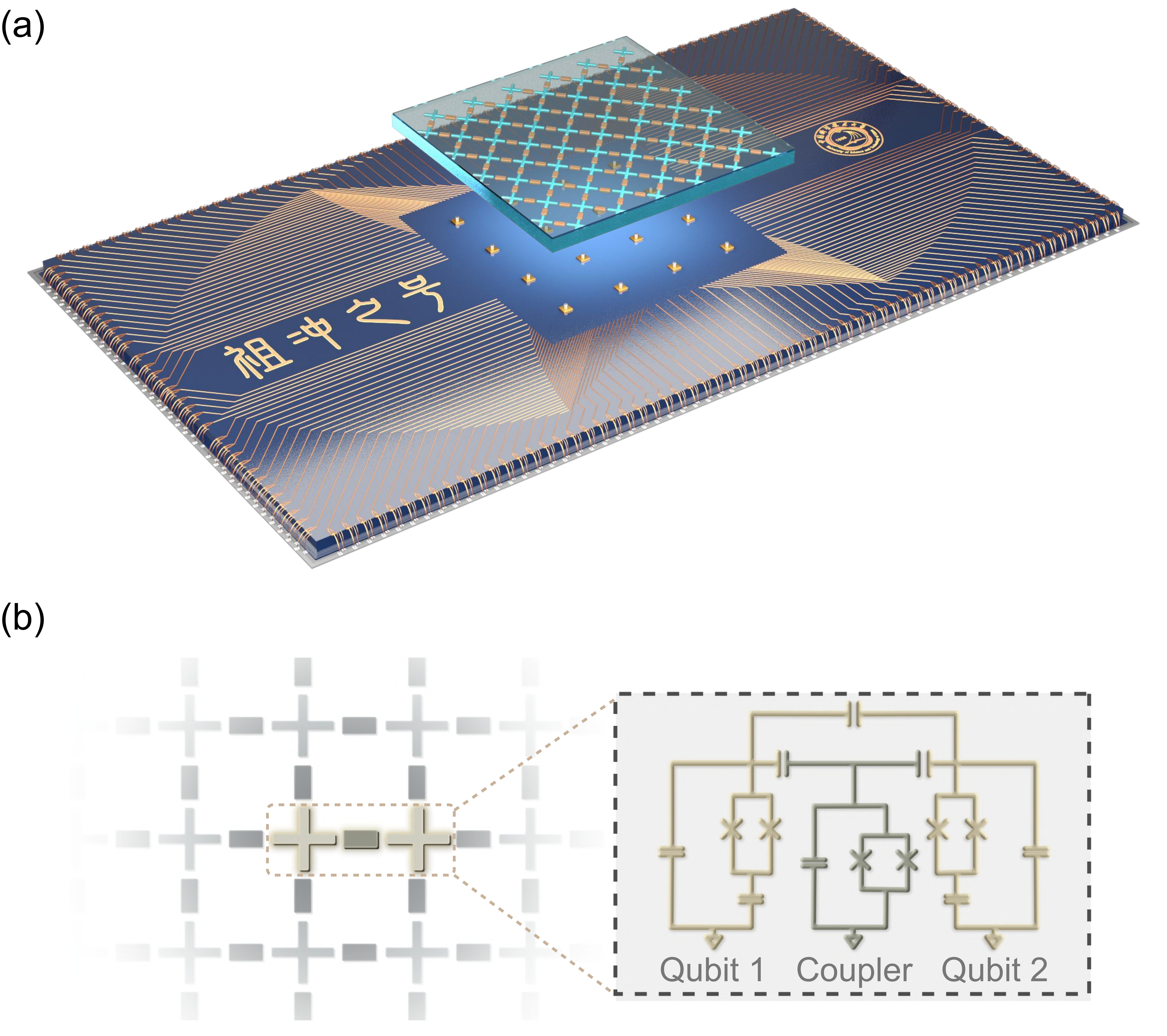}
\end{center}
\caption{\textbf{Device schematic of the \textit{Zuchongzhi} quantum processor.} (a) The \textit{Zuchongzhi} quantum processor consists of two saphire chips. One carries 66 qubits and 110 couplers, and each qubit couples to four neighboring qubits except those at the boundaries. The other hosts the readout components and control lines as well as wiring. These two chips are aligned and bounded together with indium bumps. (b) Simplified circuit schematic of the qubit and coupler.
\label{fig1}}
\end{figure}

%As a first step towards fault tolerant quantum computing, we designed a prototype quantum computer system named \textit{Zuchongzhi} which features a high fidelity quantum processor with a scalable architecture that is compatible with surface code error correction\cite{scec}.
The \textit{Zuchongzhi} quantum processor consists of 66 qubits, arrayed in 11 rows and 6 columns forming a two dimensional rectangular lattice pattern as depicted in the device schematic in Fig.~\ref{fig1}(a). The quantum processor uses Transmon qubits~\cite{koch:042319}, which are essentially non-linear oscillators with their non-linearity originating from superconducting Josephson effect. The lowest two energy levels of the non-linear oscillator are singled out to form the computational space of a qubit, encoded as $\vert0\rangle$ and $\vert1\rangle$. Each qubit has two control lines to provide full control of the qubit: a microwave drive line to drive excitations between $\vert0\rangle$ and $\vert1\rangle$, and a magnetic flux bias line to tune the qubit resonance frequency. Each qubit, except those at the boundaries, has four tunable couplers to couple to its nearest neighbors~\cite{yysCouplerCZ}, with tunable coupling that can be turned on and off with fast control.
%To implement fast, high fidelity two qubit operations, a tunable coupler~\cite{coupler} is introduced between each pair of neighboring qubits, the data qubits.
% yysCouplerCZ
% ~\cite{PhysRevLett.113.220502, yan2018tunable}
The tunable couplers are also Transmon qubits (Fig.~\ref{fig1}(b)), with frequencies several GHz higher than that of the data qubits and always stays at ground states ~\cite{yan2018tunable}. A magnetic flux bias line is provided for each coupler to fast tune the coupling strength $g$ between neighboring qubits continually from $\sim+5~\text{MHz}$ to $\sim-50~\text{MHz}$.
%, enabling fast, high fidelity two-qubit gates.
Each qubit dispersively couples to a readout resonator which  couples to a Purcell filter shared between six qubits, frequency multiplexing~\cite{ISI:000243867300038, PhysRevLett.112.190504} is used to readout the qubit states simultaneously.

All the quantum circuit components of our quantum processor are fabricated on two separate sapphire chips, which are then stacked together with the indium bump flip-chip technique. The quantum processor chip is wire bounded to a circuit board, mounted into a well shielded cryostat, and connected to room temperature control electronics through various microwave components in the wiring.

%Each qubit can be controlled independently and is coupled to its nearest neighbors with tunable coupling that can be turned on and off with fast control, enabling fast, high fidelity single qubit and two qubit quantum operations.

%Our quantum processor is a superconducting quantum integrated circuit (QIC) chip, fabricated with technologies from the semiconductor integrated circuit industry, adapted for aluminum based circuitry and superconducting Josephson junctions~\cite{JJ}. When cooled down to below 20mK, the aluminum metallic circuit become superconducting and the electrons condensed into a macroscopic state, the circuit can be considered as a macroscopic quantum system and can be engineered to perform quantum computing and quantum simulation tasks.

All the 66 qubits and 110 couplers on the quantum processor function properly. Rough calibration results for all these 66 qubits, including their decoherence time $T_1$ (average 30.6$\mu$s at idle frequencies), single-qubit gate (average 99.86\%), two-qubit gate (average 99.24\%), readout (average 95.23\%), are provided in the Supplemental Material. In this work, we select 56 qubits to demonstrate the random circuit sampling, which are optimized to achieve an optimal computational complexity in the classical simulation.
%To ensure high computational complexity in classical simulation, 56 qubits are selected for achieving the sampling task of random quantum circuits with sufficiently deep circuits. Next we will provide the result of fine-tuning on these 56 qubits.

\begin{figure*}[!htbp]
\begin{center}
\includegraphics[width=\linewidth]{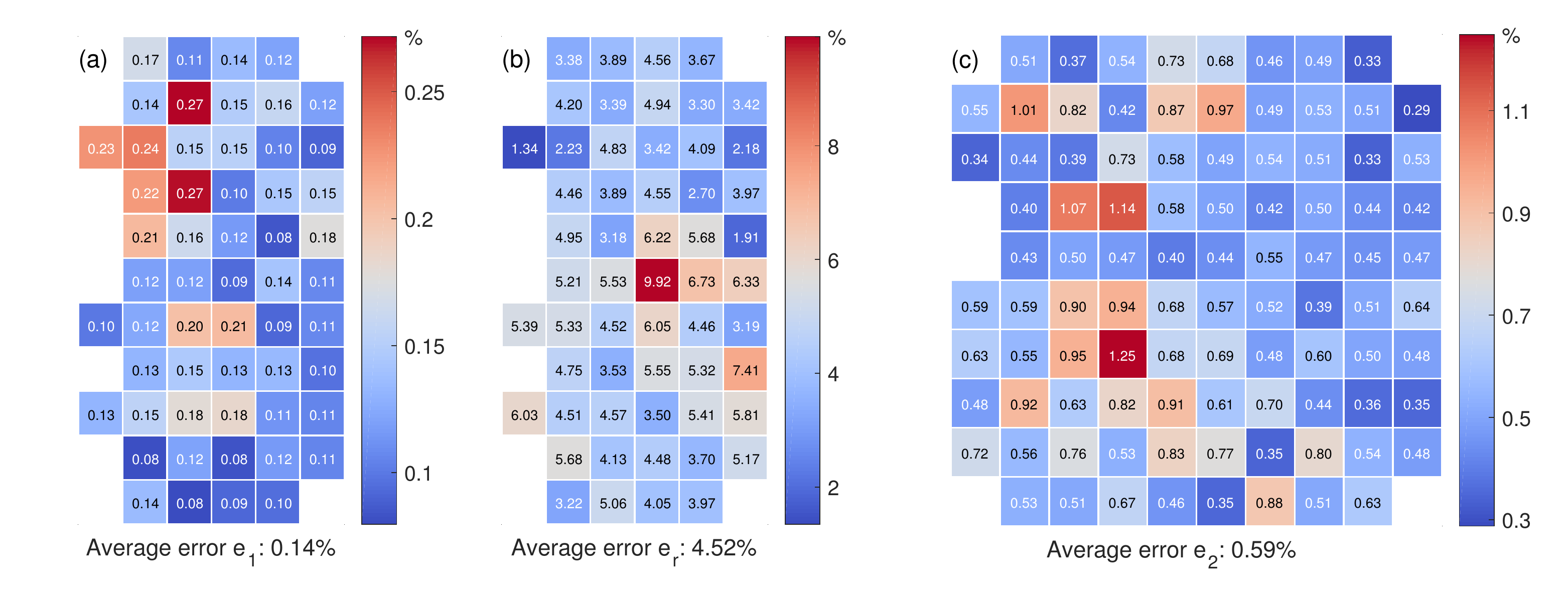}
\end{center}
\caption{\textbf{Single-qubit gate, two-qubit gate and readout performance of the selected 56 qubits.} Single-qubit gate pauli error $e_{1}$ (a), qubit state readout error $e_{r}$ (b) and two qubit gate pauli error $e_{e}$ (c) of the 56 qubits and the 94 couplers used in the random circuit sampling task. The values are provided for all qubits operating simultaneously. See Supplemental Material for the rough calibration results of all 66 qubits and 110 couplers.
\label{fig2}}
\end{figure*}

We start by calibrating the single-qubit gates. Single-qubit gates are implemented with radio-frequency (RF) pulses as the qubit frequencies are in the range of 4-6 GHz. Coherent RF pulses resonant with the qubit frequency are fed to the qubits through the microwave control lines to excite the qubits. Pulse shaping is calibrated to prevent leakage outside of the computational space~\cite{PhysRevLett.103.110501}. To enable parallel execution of gates, all the couplers are turned off when single-qubit gates are applied to isolate each qubit.
Single-qubit gate performance is susceptible to a number of conditions like coupling to two-level system (TLS), coupling to microwave resonance, microwave crosstalk and residual coupling between qubits. These conditions are mostly qubit frequency dependent, we use an error model to account for a bucket of gate error sources and learn an optimal qubit frequency configuration for all qubits through an optimization process. With the optimal qubit frequency configuration, we are able to obtain high performance single-qubit gates for all qubits. We use parallel cross-entropy benchmarking (XEB)~\cite{boixo2018characterizing, Neill195} to benchmark single-qubit gate performance. Results show an average single-qubit gate pauli error $e_1$ of $0.14\%$ when gates are applied simultaneously (Fig.~\ref{fig2}(a)).

For the random circuit sampling task, iSWAP-like gate~\cite{arute2019quantum} is used as the two-qubit gate. We bias neighboring qubits into resonance and turn on a coupling of $g\sim 10~\text{MHz}$ for a time duration $\sim 32~\rm{\text{ns}}$, which introduces swap between the qubits, as well as controlled phase interaction and single qubit phase accumulations. All these effects can be modeled as the following unitary~\cite{arute2019quantum}:
	\begin{equation}
		\begin{bmatrix}
			1 	& 0 											& 0												& 0 \\
			0 	& e^{i(\Delta_{+}+\Delta_{-})}\rm{cos}\theta 		& -ie^{i(\Delta_{+}-\Delta_{-,\text{off}})}\rm{sin}\theta	& 0 \\
			0 	& -ie^{i(\Delta_{+}+\Delta_{-,\text{off}})}\rm{sin}\theta	& e^{i(\Delta_{+}-\Delta_{-})}\rm{cos}\theta			& 0 \\
			0 	& 0 											& 0												& e^{i(2\Delta_{+}-\phi)}
		\end{bmatrix}
	\end{equation}
Parallel XEB is also employed to benchmark the iSWAP-like gate performance, an optimization process is used to learn the five parameters $\theta, \phi, \Delta_{+}, \Delta_{-}$ and $\Delta_{-,\text{off}}$ by maximizing the XEB fidelities. The length of the flux bias pulses are chosen to minimize leakage to higher energy levels, pulse distortion and timing are carefully calibrated. The qubit frequencies at which two-qubit gates are performed are also optimized following a similar procedure as setting the single-qubit operation frequencies to mitigate the influences of TLS, crosstalk and pulse distortion on gate performance. Average two-qubit gate pauli error $e_2$ of our processor is $0.59\%$ when all gates are applied simultaneously (Fig.~\ref{fig2}(c)).

To optimize readout fidelity and reduce readout crosstalk, a different frequency setting for the qubits and couplers is used when performing readout. We calibrate the readout fidelities by preparing all qubits at $|0\rangle(|1\rangle)$ and count the events of successfully identify the readout result as $|0\rangle(|1\rangle)$. The average single-qubit state readout error of our processor is $4.52\%$ (Fig.~\ref{fig2}(b)). We also compare the fidelity results with that obtained from preparing the qubits in random bit strings as a sanity check, see Suplementary Material for details.
\section{Random Quantum Circuit Benchmarking}

\begin{figure}[!tbp]
\begin{center}
\includegraphics[width=\linewidth]{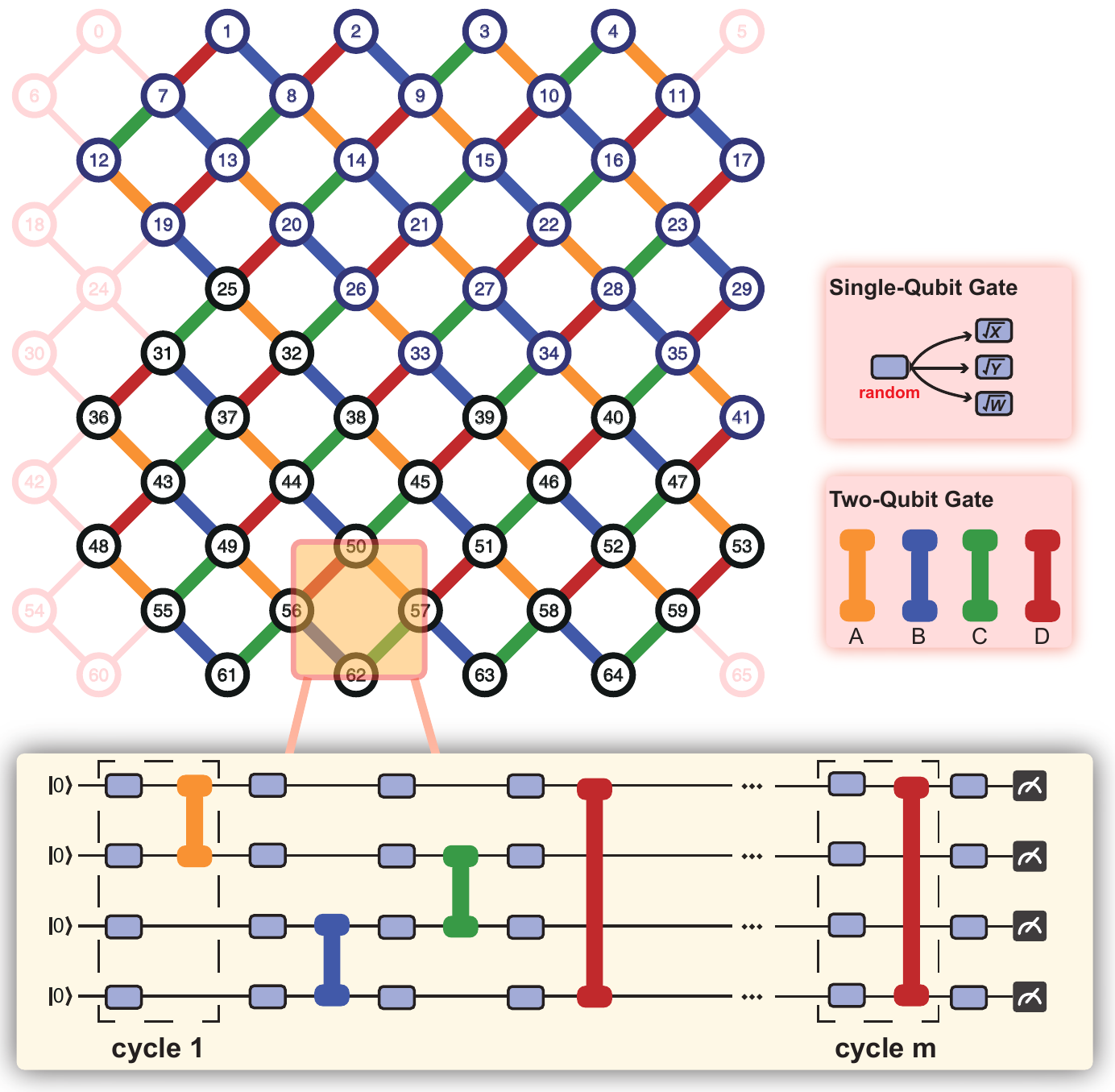}
\end{center}
\caption{\textbf{56-qubit random quantum circuit operations.} The circuit can be divided into $m$ cycles, and each cycle has a layer of single-qubit gates and two-qubit gates. The single-qubit gates are chosen randomly from the set of $\{ \sqrt X ,\sqrt Y ,\sqrt W \}$, while the two-qubit gates are chosen from the patterns of A, B, C, and D in the sequence of ABCDCDAB. The circles in the upper left corner of the diagram represent qubits, and the discarded qubits are marked with a shaded colour. The orange, blue, green, and red lines represent the two-qubit gates of the four patterns A, B, C, and D respectively.
\label{main_3}}
\end{figure}

\begin{figure*}
 \centering
\includegraphics[width=\linewidth]{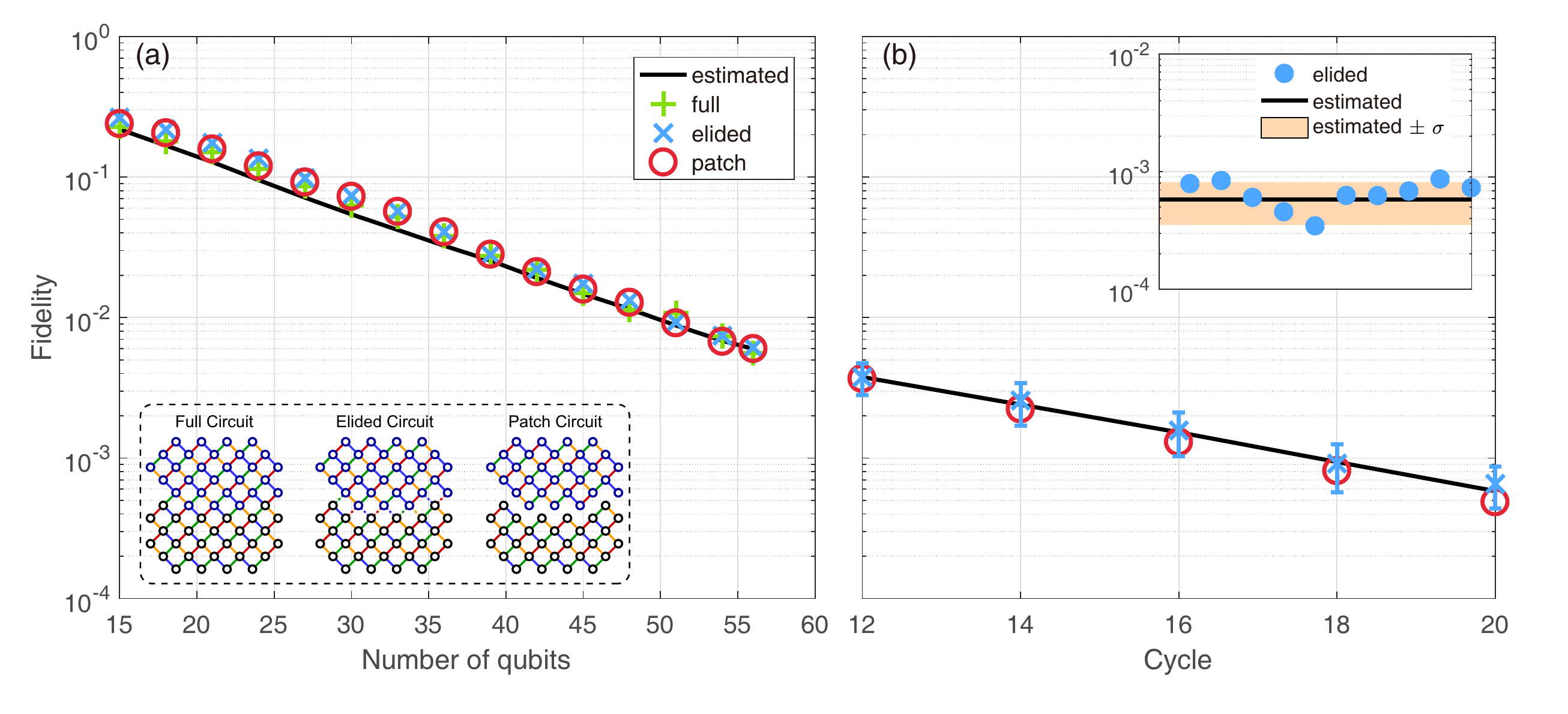}
 \caption{\textbf{Experimental results of random quantum circuits.} (a) Results of random quantum circuits with 15-56 qubits and 10 cycles. Each data point, including the results from full circuit, patch circuit, and elided circuit, is an average over six quantum circuit instances. The predicted fidelity result is shown as a black line, which is determined by the product of three types of errors, single-qubit error, two-qubit gate error and readout error. The results from the patch and elided circuits can be in good agreement with the results of the full circuit. (b) Results of random quantum circuits with 56 qubits and 12-20 cycles. For each cycle, we have sampled ten distinct random quantum circuit instances for patch, elided and full circuits. We calculate the average fidelity of the patch and elided circuits as an estimation of the fidelity of the full circuit. The error bar denotes
three standard deviations.
%, which \hhl{includes both systematic and statistical uncertainties.}
It cost about 230s to sampling 1 million bitstrings. For each 56-qubit 20-cycle circuit instance, about 19 million bitstrings are sampled in 1.2 hours and the XEB fidelity is shown in the inset. The averaged XEB fidelity of 56-qubit 20-cycle circuit over ten instances is $(6.62\pm 0.72)\times 10^{-4}$.
\label{fig4}}
\end{figure*}

To characterize the overall performance of the quantum processor, we employ the task of random quantum circuit sampling for benchmarking. Random quantum circuit is outstanding candidate to demonstrate quantum computational advantages,
%of near-term quantum computers over their conventional counterparts
and has potential applications in certified random bits~\cite{aaronson2018certified}, error correction~\cite{gullans2020quantum}, and hydrodynamics simulation~\cite{richter2020simulating}.

Figure~\ref{main_3} shows the gate sequence of our random quantum circuit. Each random quantum circuit is composed of $m$ cycles, and each cycle is composed of a single-qubit gate layer and a two-qubit gate layer. In the single-qubit gate layer, single-qubit gates are applied on all qubits and chosen randomly from the set of $\{ \sqrt X ,\sqrt Y ,\sqrt W \}$, where $\sqrt X {\rm{ = }}{R_X}(\pi /2)$, $\sqrt Y {\rm{ = }}{R_Y}(\pi /2)$, and $\sqrt W {\rm{ = }}{R_{(X+Y)}}(\pi /2)$ are $\pi /2$-rotation around
specific axis. Each single-qubit gate on a qubit in subsequent cycle is independently and randomly chosen from the subset of $\{ \sqrt X ,\sqrt Y ,\sqrt W \}$, which does not include the single-qubit gate to this qubit in the preceding cycle. In the two-qubit gate layer, two-qubit gates are applied according to a specified pattern, labeled by A, B, C, and D, in sequence of ABCDCDAB. Finally, an additional single-qubit gate layer is applied after $m$ cycles and before measurement.

With just a few cycles, the random quantum circuit could generate a highly entangled state.
%, which is extremely difficult to reproduce with even the best supercomputers. Thus,
Two variant circuits, patch circuit and elided circuit, are utilized to estimate the XEB fidelity of quantum circuits within our classical computing capabilities. The ``patch" circuits are designed by removing a slice of two-qubit gates, while the ``elided" circuits only remove a fraction of the gates between the patches. In this two variant circuits, the amount of entanglement involved is reduced so that it is feasible to classically simulate the experiments and thus determine $F_{\text{XEB}}$. We test the linear XEB fidelities of these two variant circuits and full version of the circuits ranging from 15 qubits to 56 qubits with 10 cycles (see Fig.~\ref{fig4}(a)). Over all of these circuits, the fidelities derived from patch and elided circuits are in good agreement with the fidelities obtained with the corresponding full circuits, with average deviations of $\sim5\%$ and $\sim10\%$, respectively, dominated by system fluctuations. The achieved results indicate that patch circuits and elided circuits could used as performance estimators for large systems.

%From the experimental results, we find that the XEB fidelities of full circuits, patch circuits, and elided circuits all remain in agreement, indicating that patch circuits and elided circuits could used as performance estimators for large systems.

We now turn to test 56-qubit circuits increasingly with more cycles. The output bitstrings of full, patch, and elided circuits from 12 to 20 cycles are all sampled in our experiments. However, the verification of full circuit becomes challenging in this regime due to our limited classical computing resources. Therefore, we use the previously tested patch and elided circuits to assess performance. Figure.~\ref{fig4}(b) shows the linear XEB results for patch circuits and elided circuits. For each cycle, a total of ten randomly generated circuit instances are executed and sampled. We collect approximately $1.9 \times {10^7}$ bitstrings for each 56-qubit circuit with 20 cycles, the fidelities for these ten elided circuits are given in the inset of Fig.~\ref{fig4}(b). Each individual circuit instance fidelity is nearly inside the $\pm \sigma$ statistical error band for a single instance, indicating the stability of the system and the unbiasedness of noise. We then apply inverse-variance weighting over these ten random circuits, yielding $F=(6.62\pm 0.72)\times 10^{-4}$ for the combined linear XEB fidelity of the 56-qubit 20-cycle circuit. The null hypothesis of uniform sampling ($F=0$) is thus rejected with a significance of $9\sigma$.

In addition, the observed fidelity of each circuit, as well as the decay of XEB fidelities with qubits $n$ and cycles $m$, match the predicted fidelity calculated from a simple multiplication of individual operations quite well.
%~(e.g. the predicted fidelity of 56-qubit 20-cycle circuit is $5.766\times 10^{-4}$, and there is only a 2.5\% deviation between the measured and predicted fidelities.).
This result provides convincing evidence to confirm the low correlation of errors of each individual operation, including single- and two-qubit gates, as well as readout, which is a critical aspect for quantum error correction.

\section{Computational Cost Estimation}

We finally estimate the classical computational cost of our hardest circuits, i.e. 56-qubit random quantum circuit with 20 cycles. The estimation is based on two types of classical algorithms which are considered state-of-the-art for classically simulating quantum circuits, namely tensor network algorithm and the Schr$\ddot{\text{o}}$dinger-Feynman algorithm.

Tensor network algorithm reduces the problem of computing amplitudes into contracting tensor networks. It is a single-amplitude algorithm in that the complexity grows linearly with the number of amplitudes, which has been shown to perform excelently for relatively shallow quantum circuits~\cite{MarkovShi2008,GuoWu2019,VillalongaMandra2019,villalonga2020establishing,HuangChen2020,pan2021simulating,guo2021verifying}. The computational cost of tensor networks algorithms is determined by the tensor contract path. To identify an optimal tensor contract path, we use the python package cotengra~\cite{cotengra}, which has been shown to be capable of reproducing state-of-the-art results in Ref.~\cite{HuangChen2020,pan2021simulating}. The number of floating point operations to generate one perfect sample from the $53$-qubit 20-cycle random circuit used in Ref.~\cite{arute2019quantum} and our 56-qubit and 20-cycle random circuit are thus estimated as $1.63\times 10^{18}$ and $1.65\times 10^{20}$, respectively. Given that $3 \times 10^6$ samples were collected over one circuit instance with $0.224\%$ fidelity in Ref.~\cite{arute2019quantum}, while we have collected $1.9 \times 10^7$ samples with $0.0662\%$ fidelity, so theoretically it would cost a total of $1.10\times 10^{22}$ and $2.08\times 10^{24}$ floating point operations, respectively, to reproduce the same results as Ref.~\cite{arute2019quantum} and our work using classical computer (see Supplemental Material for details).
%, and a discussion about the subspace sampling trick in Ref.~\cite{pan2021simulating}).
%We note that in Ref.~\cite{pan2021simulating} a subspace sampling trick is used, which computes all the amplitudes in one go for a subspace of $21$ qubits, with the rest qubits fixed to $0$s. The number of floating point operations for the $53$-qubit random circuit with $20$ cycles used in Ref.~\cite{arute2019quantum} is thus estimated as $1.63\times 10^{18}$.  To estimate the complexity of our problem we follow this approach by choosing a subspace of $25$ qubits (therefore we will get about $30,000,000$ amplitudes which roughly matches the number we used in our experiment), and get a number of $1.42\times 10^{20}$ floating point operations (see Supplemental Material for details).

In comparison, the Schr$\ddot{\text{o}}$dinger-Feynman algorithm is a full-amplitude algorithm in that computing an arbitrarily chosen branch of amplitudes is almost as hard as computing a single amplitude. Similar to Ref.~\cite{arute2019quantum}, we estimate that it would cost $5.76\times 10^{17}$ core-hours for the task of simulating 56-qubtit 20-cycle random quantum circuit sampling with $0.0662\%$ fidelity using the Schr$\ddot{\text{o}}$dinger-Feynman algorithm, while simulating the previous task on the 53-qubit 20-cycle circuit ($0.224\%$ fidelity~\cite{arute2019quantum}) would cost $8.90\times 10^{13}$ core-hours (see Supplemental Material for details).

Therefore, using the tensor network algorithm or Schr$\ddot{\text{o}}$dinger-Feynman algorithm, the classical computational cost of our sample task with 56-qubit and 20-cycle is about 2-3 orders of magnitude greater than that of the previous task with 53-qubit and 20-cycle~\cite{arute2019quantum}. This indicates that our work significantly enlarges the gap between the computational advantages of quantum devices and the classical simulations.
In particular, as discussed in the Supplemental Material, it is estimated that it will take 15.9 days to simulate the previous sampling task in Ref.~\cite{arute2019quantum} using tensor network algorithm on Summit, whereas simulating our sampling task will take 8.2 years.
We anticipate the development of more efficient classical simulation approaches. On the one hand, the competition between
%of performance comparisons between
quantum and classical computing will continue; on the other hand, more efficient classical simulation methods are necessary for large-scale quantum computing benchmarking.

%The Schr$\ddot{\text{o}}$dinger-Feynman algorithm separates the circuit into several groups of two patches of qubits, and then utilizes Feynman path-integral approach to gather all the outcomes of these groups simulated by Schr$\ddot{\text{o}}$dinger method.

\section{Conclusion} \label{sec:summary}

In conclusion, we have reported the design, fabrication, measurement, and benchmarking of a state-of-the-art 66-qubit superconducting quantum processor that is fully programmable through electric control. We are able to achieve high-fidelity logic operations of the full quantum circuit and eliminate the unwanted cross talk. Our experimental results of random quantum circuit with 56 qubits and 20 cycles on \textit{Zuchongzhi} quantum processor established a new record to challenge the classical computing capability. We note that the performance of the whole system behaves as predicted when system size grows from small to large, confirming our high-fidelity quantum operations and low correlated errors on the \textit{Zuchongzhi} processor. The quantum processor has a scalable architecture that is compatible with surface-code error correction, which can act as the test-bed for fault-tolerant quantum computing. We also expect this large-scale, high-performance quantum processor could enable us to pursue valuable NISQ quantum applications beyond classical computers in the near future.

\begin{acknowledgments}
We thank Run-Ze Liu, Wen Liu, Chenggang Zhou, Pan Zhang, Junjie Wu for very helpful discussions and assistance. The classical calculations were performed on the supercomputing system in the Supercomputing Center of University of Science and Technology of China. The authors thank the USTC Center for Micro- and Nanoscale Research and Fabrication for supporting the sample fabrication. The authors also thank QuantumCTek Co., Ltd., for supporting the fabrication and the maintenance of room-temperature electronics.
 \textbf{Funding:}
 This research was supported by the National Key R\&D Program of China (Grant No. 2017YFA0304300), the Chinese Academy of Sciences, Anhui Initiative in Quantum Information Technologies, Technology Committee of Shanghai Municipality, National Natural Science Foundation of China (Grants No. 11905217, No. 11774326, Grants No. 11905294), Shanghai Municipal Science and Technology Major Project (Grant No. 2019SHZDZX01), Natural Science Foundation of Shanghai (Grant No. 19ZR1462700), Key-Area Research and Development Program of Guangdong Provice (Grant No. 2020B0303030001), and the Youth Talent Lifting Project (Grant No. 2020-JCJQ-QT-030).

The authors' names appear in alphabetical order by last name.
\end{acknowledgments}

\bibliographystyle{apsrev4-1}
\bibliography{references}

\end{document}

\title{Supplemental Material for \\ ``Strong quantum computational advantage using a superconducting quantum processor''}

\author{Yulin Wu}
\affiliation{Hefei National Laboratory for Physical Sciences at the Microscale and Department of Modern Physics, University of Science and Technology of China, Hefei 230026, China}
\affiliation{Shanghai Branch, CAS Center for Excellence in Quantum Information and Quantum Physics, University of Science and Technology of China, Shanghai 201315, China}
\affiliation{Shanghai Research Center for Quantum Sciences, Shanghai 201315, China}
\author{Wan-Su Bao}
\affiliation{Henan Key Laboratory of Quantum Information and Cryptography, Zhengzhou 450000, China}
\author{Sirui Cao}
\author{Fusheng Chen}
\author{Ming-Cheng Chen}
\affiliation{Hefei National Laboratory for Physical Sciences at the Microscale and Department of Modern Physics, University of Science and Technology of China, Hefei 230026, China}
\affiliation{Shanghai Branch, CAS Center for Excellence in Quantum Information and Quantum Physics, University of Science and Technology of China, Shanghai 201315, China}
\affiliation{Shanghai Research Center for Quantum Sciences, Shanghai 201315, China}
\author{Xiawei Chen}
\affiliation{Shanghai Branch, CAS Center for Excellence in Quantum Information and Quantum Physics, University of Science and Technology of China, Shanghai 201315, China}
\author{Tung-Hsun Chung}
\author{Hui Deng}
\affiliation{Hefei National Laboratory for Physical Sciences at the Microscale and Department of Modern Physics, University of Science and Technology of China, Hefei 230026, China}
\affiliation{Shanghai Branch, CAS Center for Excellence in Quantum Information and Quantum Physics, University of Science and Technology of China, Shanghai 201315, China}
\affiliation{Shanghai Research Center for Quantum Sciences, Shanghai 201315, China}
\author{Yajie Du}
\affiliation{Shanghai Branch, CAS Center for Excellence in Quantum Information and Quantum Physics, University of Science and Technology of China, Shanghai 201315, China}
\author{Daojin Fan}
\author{Ming Gong}
\author{Cheng Guo}
\author{Chu Guo}
\author{Shaojun Guo}
\author{Lianchen Han}
\affiliation{Hefei National Laboratory for Physical Sciences at the Microscale and Department of Modern Physics, University of Science and Technology of China, Hefei 230026, China}
\affiliation{Shanghai Branch, CAS Center for Excellence in Quantum Information and Quantum Physics, University of Science and Technology of China, Shanghai 201315, China}
\affiliation{Shanghai Research Center for Quantum Sciences, Shanghai 201315, China}
\author{Linyin Hong}
\affiliation{QuantumCTek Co., Ltd., Hefei 230026, China}
\author{He-Liang Huang}
\affiliation{Hefei National Laboratory for Physical Sciences at the Microscale and Department of Modern Physics, University of Science and Technology of China, Hefei 230026, China}
\affiliation{Shanghai Branch, CAS Center for Excellence in Quantum Information and Quantum Physics, University of Science and Technology of China, Shanghai 201315, China}
\affiliation{Shanghai Research Center for Quantum Sciences, Shanghai 201315, China}
\affiliation{Henan Key Laboratory of Quantum Information and Cryptography, Zhengzhou 450000, China}
\author{Yong-Heng Huo}
\affiliation{Hefei National Laboratory for Physical Sciences at the Microscale and Department of Modern Physics, University of Science and Technology of China, Hefei 230026, China}
\affiliation{Shanghai Branch, CAS Center for Excellence in Quantum Information and Quantum Physics, University of Science and Technology of China, Shanghai 201315, China}
\affiliation{Shanghai Research Center for Quantum Sciences, Shanghai 201315, China}
\author{Liping Li}
\affiliation{Shanghai Branch, CAS Center for Excellence in Quantum Information and Quantum Physics, University of Science and Technology of China, Shanghai 201315, China}
\author{Na Li}
\author{Shaowei Li}
\author{Yuan Li}
\author{Futian Liang}
\affiliation{Hefei National Laboratory for Physical Sciences at the Microscale and Department of Modern Physics, University of Science and Technology of China, Hefei 230026, China}
\affiliation{Shanghai Branch, CAS Center for Excellence in Quantum Information and Quantum Physics, University of Science and Technology of China, Shanghai 201315, China}
\affiliation{Shanghai Research Center for Quantum Sciences, Shanghai 201315, China}
\author{Chun Lin}
\affiliation{Shanghai Institute of Technical Physics, Chinese Academy of Sciences, Shanghai 200083, China}
\author{Jin Lin}
\author{Haoran Qian}
\affiliation{Hefei National Laboratory for Physical Sciences at the Microscale and Department of Modern Physics, University of Science and Technology of China, Hefei 230026, China}
\affiliation{Shanghai Branch, CAS Center for Excellence in Quantum Information and Quantum Physics, University of Science and Technology of China, Shanghai 201315, China}
\affiliation{Shanghai Research Center for Quantum Sciences, Shanghai 201315, China}
\author{Dan Qiao}
\affiliation{Shanghai Branch, CAS Center for Excellence in Quantum Information and Quantum Physics, University of Science and Technology of China, Shanghai 201315, China}
\author{Hao Rong}
\author{Hong Su}
\author{Lihua Sun}
\affiliation{Hefei National Laboratory for Physical Sciences at the Microscale and Department of Modern Physics, University of Science and Technology of China, Hefei 230026, China}
\affiliation{Shanghai Branch, CAS Center for Excellence in Quantum Information and Quantum Physics, University of Science and Technology of China, Shanghai 201315, China}
\affiliation{Shanghai Research Center for Quantum Sciences, Shanghai 201315, China}
\author{Liangyuan Wang}
\affiliation{Shanghai Branch, CAS Center for Excellence in Quantum Information and Quantum Physics, University of Science and Technology of China, Shanghai 201315, China}
\author{Shiyu Wang}
\author{Dachao Wu}
\author{Yu Xu}
\affiliation{Hefei National Laboratory for Physical Sciences at the Microscale and Department of Modern Physics, University of Science and Technology of China, Hefei 230026, China}
\affiliation{Shanghai Branch, CAS Center for Excellence in Quantum Information and Quantum Physics, University of Science and Technology of China, Shanghai 201315, China}
\affiliation{Shanghai Research Center for Quantum Sciences, Shanghai 201315, China}
\author{Kai Yan}
\affiliation{Shanghai Branch, CAS Center for Excellence in Quantum Information and Quantum Physics, University of Science and Technology of China, Shanghai 201315, China}
\author{Weifeng Yang}
\affiliation{QuantumCTek Co., Ltd., Hefei 230026, China}
\author{Yang Yang}
\affiliation{Shanghai Branch, CAS Center for Excellence in Quantum Information and Quantum Physics, University of Science and Technology of China, Shanghai 201315, China}
\author{Yangsen Ye}
\affiliation{Hefei National Laboratory for Physical Sciences at the Microscale and Department of Modern Physics, University of Science and Technology of China, Hefei 230026, China}
\affiliation{Shanghai Branch, CAS Center for Excellence in Quantum Information and Quantum Physics, University of Science and Technology of China, Shanghai 201315, China}
\affiliation{Shanghai Research Center for Quantum Sciences, Shanghai 201315, China}
\author{Jianghan Yin}
\affiliation{Shanghai Branch, CAS Center for Excellence in Quantum Information and Quantum Physics, University of Science and Technology of China, Shanghai 201315, China}
\author{Chong Ying}
\author{Jiale Yu}
\author{Chen Zha}
\author{Cha Zhang}
\affiliation{Hefei National Laboratory for Physical Sciences at the Microscale and Department of Modern Physics, University of Science and Technology of China, Hefei 230026, China}
\affiliation{Shanghai Branch, CAS Center for Excellence in Quantum Information and Quantum Physics, University of Science and Technology of China, Shanghai 201315, China}
\affiliation{Shanghai Research Center for Quantum Sciences, Shanghai 201315, China}
\author{Haibin Zhang}
\affiliation{Shanghai Branch, CAS Center for Excellence in Quantum Information and Quantum Physics, University of Science and Technology of China, Shanghai 201315, China}
\author{Kaili Zhang}
\author{Yiming Zhang}
\affiliation{Hefei National Laboratory for Physical Sciences at the Microscale and Department of Modern Physics, University of Science and Technology of China, Hefei 230026, China}
\affiliation{Shanghai Branch, CAS Center for Excellence in Quantum Information and Quantum Physics, University of Science and Technology of China, Shanghai 201315, China}
\affiliation{Shanghai Research Center for Quantum Sciences, Shanghai 201315, China}
\author{Han Zhao}
\affiliation{Shanghai Branch, CAS Center for Excellence in Quantum Information and Quantum Physics, University of Science and Technology of China, Shanghai 201315, China}
\author{Youwei Zhao}
\affiliation{Hefei National Laboratory for Physical Sciences at the Microscale and Department of Modern Physics, University of Science and Technology of China, Hefei 230026, China}
\affiliation{Shanghai Branch, CAS Center for Excellence in Quantum Information and Quantum Physics, University of Science and Technology of China, Shanghai 201315, China}
\affiliation{Shanghai Research Center for Quantum Sciences, Shanghai 201315, China}
\author{Liang Zhou}
\affiliation{QuantumCTek Co., Ltd., Hefei 230026, China}
\author{Qingling Zhu}
\author{Chao-Yang Lu}
\author{Cheng-Zhi Peng}
\author{Xiaobo Zhu}
\author{Jian-Wei Pan}

\affiliation{Hefei National Laboratory for Physical Sciences at the Microscale and Department of Modern Physics, University of Science and Technology of China, Hefei 230026, China}
\affiliation{Shanghai Branch, CAS Center for Excellence in Quantum Information and Quantum Physics, University of Science and Technology of China, Shanghai 201315, China}
\affiliation{Shanghai Research Center for Quantum Sciences, Shanghai 201315, China}

\date{\today}

\maketitle

\setcounter{section}{0}
\renewcommand{\thefigure}{S\arabic{figure}}	
\renewcommand{\thetable}{S\arabic{table}}	
\renewcommand{\theequation}{S\arabic{equation}}	
\setcounter{figure}{0}
\setcounter{table}{0}
\setcounter{equation}{0}

\section{Quantum processor design and fabrication}
We designed a state of the art programmable quantum processor, consists of 66 Transmon~\cite{koch:042319} qubits in a two-dimensional array, with another 110 Transmon qubits as couplers for adjustable coupling between neighboring qubits, as illustrated in Fig.~1 in the main text.
The 66 qubits are divided into 11 groups, with 6 qubits in a group sharing a readout Purcell filter~\cite{PhysRevLett.112.190504}.
Taking advantage of the flip-chip technology, all qubits and couplers are placed on the same layer, and all readout and control lines are on the other layer.
Each qubit is controlled by an control line which combines both microwave drive (XY control) and flux bias (Z control), and capacitively coupled to a quarter wave resonator which is coupled to a Purcell filter for dispersive readout.
The qubit-qubit coupling of nearest-neighbor qubits are contributed by two parts: direct capacitive coupling and indirect coupling through the coupler.
Each coupler has an individual flux bias line, with which the effective qubit-qubit coupling strength can be tuned from $\sim +5 \text{MHz}$ to $\sim -50 \text{MHz}$ by changing the coupler frequency.
The two-qubit gates are implemented with a coupling strength of about $10 \text{MHz}$.

Our quantum processor consists of two separate chips, the top chip and the bottom chip. High purity aluminum thin film are grown on sapphire substrate by molecular beam epitaxy (MBE) for both chips~\cite{megrant2012planar}. Control and readout circuits are fabricated on the bottom chip using optical lithography. To suppress crosstalk, airbridges are fabricated to shield critical circuits~\cite{dunsworth2018method}. 66 qubits and 110 adjustable couplers are fabricated on the top chip using an aluminum evaporation and lift-off process. After dicing and testing, the two separated chips are bonded together with indium bumps~\cite{foxen2017qubit,rosenberg20173d}. After chip fabrication, the processor is wire-bounded to a printed circuit board inside a sample box with gold-plated shielding inside and $\mu-$metal shielding outside. Finally, the packaged processor is mounted to the cold plate of a dilution refrigerator and connected to room temperature electronics through alternators, filters and amplifiers.

\section{Experimental setup}
\begin{figure}[!htbp]
\begin{center}
\includegraphics[width=0.95\linewidth]{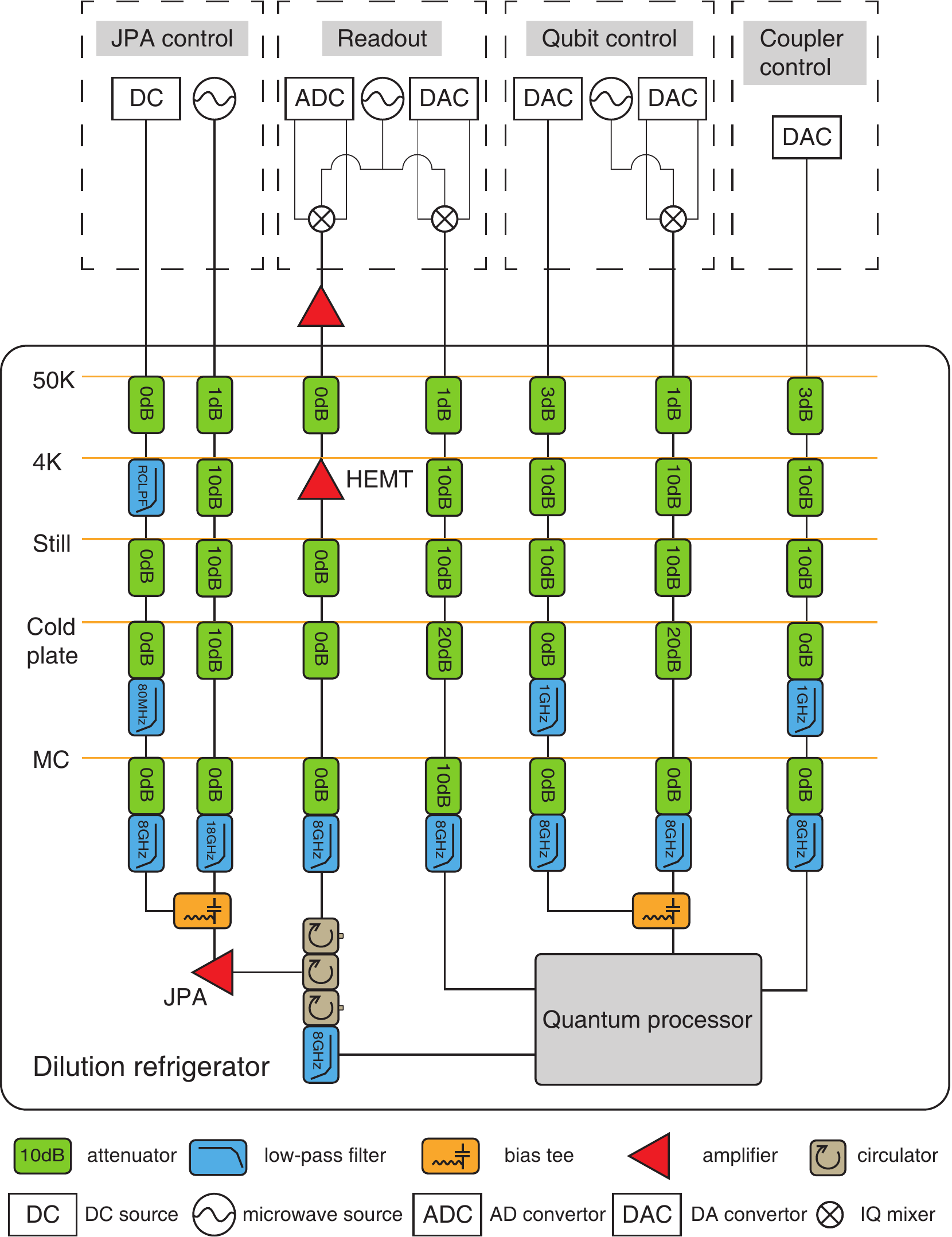}
\end{center}
\caption{\textbf{The schematic diagram of control electronics and wiring.} Each qubit has a XY control line and and a Z control line, which are combined together via a bias tee before connected to the quantum processor. In the dilution refrigerator, attenuators and filters are installed at various stages to reduce noise. Josephson parametric amplifiers (JPA) , high electron mobility transistors (HEMT) and room-temperature amplifiers are used to amplify the readout signals. At room temperature, digital-to-analog converters (DAC), microwave sources and mixers are used to generate pulses for qubit XY control, readout probing pulses and JPA pump. Qubit Z and coupler control pulses are also generated by DACs. DC sources are used to provide the flux bias of JPAs. The readout signals amplified by the room-temperature amplifiers are  digitized and demodulated by ADC modules.}

\label{fig.wiring}
\end{figure}

The experimental wiring setup for qubit/coupler controls and frequency-multiplexed readouts is shown in Fig.~\ref{fig.wiring}. The control and readout signals are generated by digital-to-analog converters (DAC) at room temperature, then attenuated by a series of attenuators and filtered by several low-pass filters.

In order to improve the signal-to-noise ratio of the readout signal, we use the Josephson parametric amplifier (JPA) for the first stage amplification at base temperature. Then
the readout signal is amplified by a high-electron mobility transistor (HEMT) amplifier at the 4 K stage and  further amplified by a room-temperature amplifiers after getting out of the dilution refrigerator.

The room temperature electronic equipment in this experiment includes 330 DAC channels, 11 ADC modules, 11 DC channels and 34 microwave source channels.

\section{Gate and Readout Calibration}
\subsection{Part1: Rough calibration on 66 qubits}
\subsubsection{Basic Calibration}
% Firstly, we do some basic calibration experiment to ensure all parts of the chip can work properly, incluing 66 qubits, 110 couplers, 66 readout resonators and 11 JPAs. The calibration procedure is as follows.
After cooling down, we perform basic calibrations to determine whether the processor can work properly, this process involves all 66 qubits, 110 couplers, 66 readout resonators and 11 JPAs. The calibration procedure is listed bellow.
\begin{itemize}

  \item Identify the the readout resonator frequency for each qubit,  measure the dispersive frequency shift and measure its dependency with the qubit flux bias.
  \item Find a JPA DC bias, pump frequency and pump power configuration that produces high signal-to-noise ratio for the corresponding readout signals.
  \item Measure the response of readout resonator versus coupler bias, then bias the coupler to a safe point where coupling strength of neighboring qubits is small, $\sim+5$MHz typically.
  \item Run Rabi experiments to calibrate $\pi$ and $\pi/2$ pulse amplitudes.
  \item Perform coarse readout  calibration to improve readout fidelity by optimizing readout frequency, length and amplitude.
  \item Tune coupler bias to minimize the qubit-qubit coupling via $|10\rangle$ - $|01\rangle$ swap experiment.
  \item Perform qubit spectroscopy measurements and extract the mapping between the qubit bias amplitude and the qubit frequency.
  \item Measure $T_1$, $T_2$ versus qubit frequency.
  \item Measure the step response of the bias control line to characterize the Z pulse distortion by running a Ramsey experiment~\cite{2019Strongly}.
  \item Perform XY-crosstalk measurement.
  \item Synchronise the timing between the qubit microwave drive, qubit bias, and coupler bias.
\end{itemize}

As qubit idle frequency arrangement and coupling between qubits are not yet optimized at this stage,  we typically divide the qubits into 2 groups and couplers into 4 groups for some calibration experiments, calibration for each group are performed in parallel. Basic calibration results indicates that the quantum processor can work properly and we proceed to optimize the frequency arrangement with the coherence and XY-crosstalk data.
There are three types of operating frequencies: idle, interaction and readout. When optimizing the frequency arrangement, we have to consider the following factors: coherence, two-level-system (TLS), residual coupling between qubits, XY-crosstalk and Z pulse distortions.

After the optimal operating frequencies are determined, we proceed to
%run fine calibration experiments to
calibrate and optimize single-qubit gate, readout and two-qubit gate parameters in parallel.
%due to small crosstalk on our chip.

\subsubsection{Readout, Single-Qubit Gate and Two-Qubit Gate Calibration}
~\\
\textbf{i. Single-Qubit Gate Calibration}

The optimized idle frequency arrangement is displayed in Fig.~\ref{66qubit}(a). We set all qubit idle frequencies to the target idle frequencies by adjusting qubit flux biases. Next, we turn off the qubit-qubit coupling by adjusting the coupler flux bias. Then we fine tune the XY drive pulse amplitudes, pulse shaping parameters of the cosine-envelop microwave of 25ns length. Finally, we benckmark the fidelities of single qubit gates with single-qubit cross-entropy benchmarking (XEB)~\cite{arute2019quantum}.

Gate errors of some qubits are still too high($\textgreater0.3\%$) for the random quantum circuit benchmarking tasks after this first round of calibration.
% We find that if a qubit is close to a TLS in frequency domain, even if this qubit have a relaxation time  longer than 20 $\mu s$, the  actual SPB error of the qubits is still much larger than that estimated throuth relaxation time.
For these qubits, we measure XEB gate fidelity versus idle frequencies around initial idle frequency and choose a new optimal idle frequency.

\begin{figure*}[!htbp]
\begin{center}
\includegraphics[width=\linewidth]{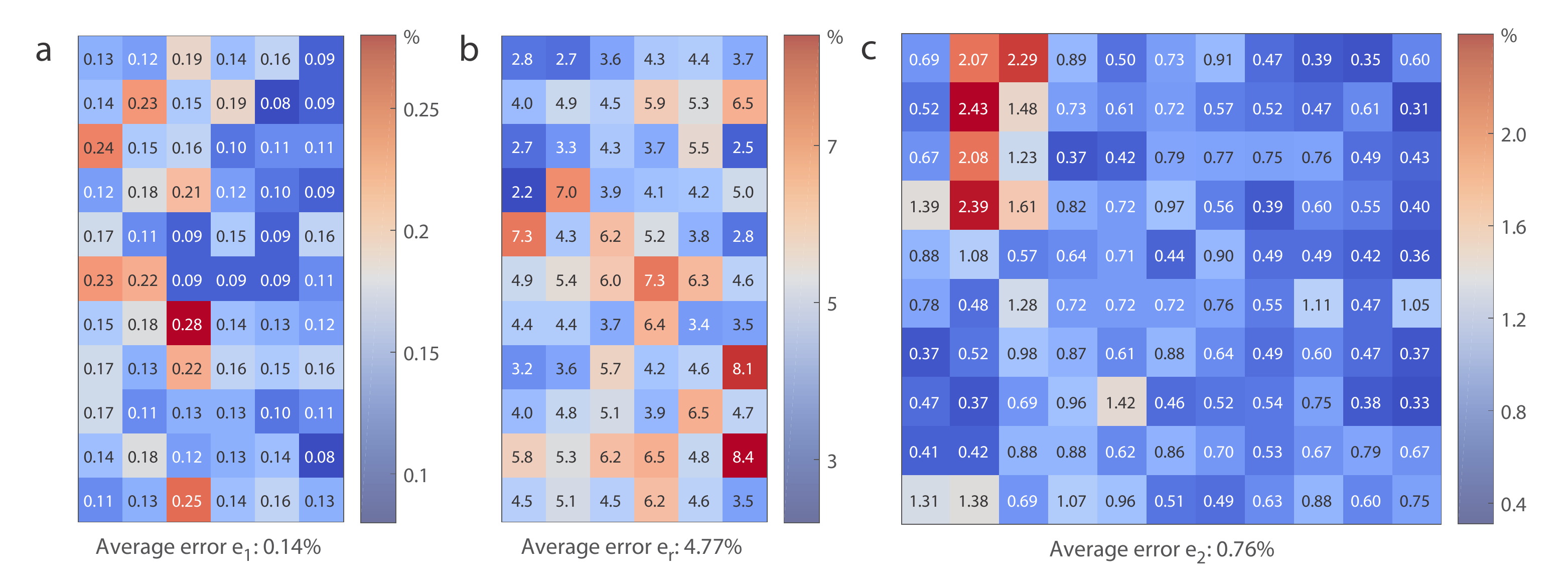}
\end{center}
\caption{\textbf{Single-qubit gate pauli error $e_1$, two-qubit gate pauli error $e_2$ and readout error $e_r$ of the 66 qubits and 110 couplers of the \textit{Zuchongzhi} processor.}
\label{66error}}
\end{figure*}

After the above calibration procedure, we obtained high fidelity single-qubit gates with an average XEB pauli error 0.14\% when applied simultaneously. Complete XEB pauli error data of each single-qubit gates are shown in Fig.~\ref{66error}(a).

~\\
\textbf{ii. Readout Calibration}

To improve readout fidelity and reduce readout crosstalk, we set the qubit readout frequencies to a different frequency arrangement optimized for readout, which is displayed in Fig.~\ref{66qubit}(e). At this new frequency setting, we rerun the basic readout calibrations to optimize readout frequency, length and amplitude.
After this calibration procedure, the readout error is mainly limited by qubit relaxation. To reduce the influence of the relaxation, we enhance the effective qubit lifetime by driving the qubit to higher levels during the readout~\cite{Elder_2020}.
% Moreover, we treat both physical $|1\rangle$ and physical $|2\rangle$ as logical $|1\rangle$ and physical $|0\rangle$ as logical $|0\rangle$. In this way, only when physical $|2\rangle$ relax to physical $|1\rangle$ and then to physical $|0\rangle$ will we mistake logical $|1\rangle$ as logical $|0\rangle$. In other words, logical $|1\rangle$ have a longer effective lifetime during measurement.

We calibrate the readout fidelities by preparing all qubits at $|0\rangle(|1\rangle)$ and count the events of successfully identify the readout result as $|0\rangle(|1\rangle)$. Simultaneous readout error of $|0\rangle$ and $|1\rangle$ are 3.46\% and 6.08\% respectively, average readout fidelity is 95.23\%, as shown in Fig.~\ref{66error}(b).

We also compare the fidelity results with that obtained from preparing the qubits at random bit strings as a sanity check. The results of random bit string measurement show that identification error of a single qubit has increased by an average of $0.14\%$ and the 56-qubit state readout fidelity is lower with a factor of $0.93$. This factor is used to correct the estimated fidelities in the random quantum circuit benchmarking tasks.

\begin{figure*}[!htbp]
\begin{center}
\includegraphics[width=0.65\linewidth]{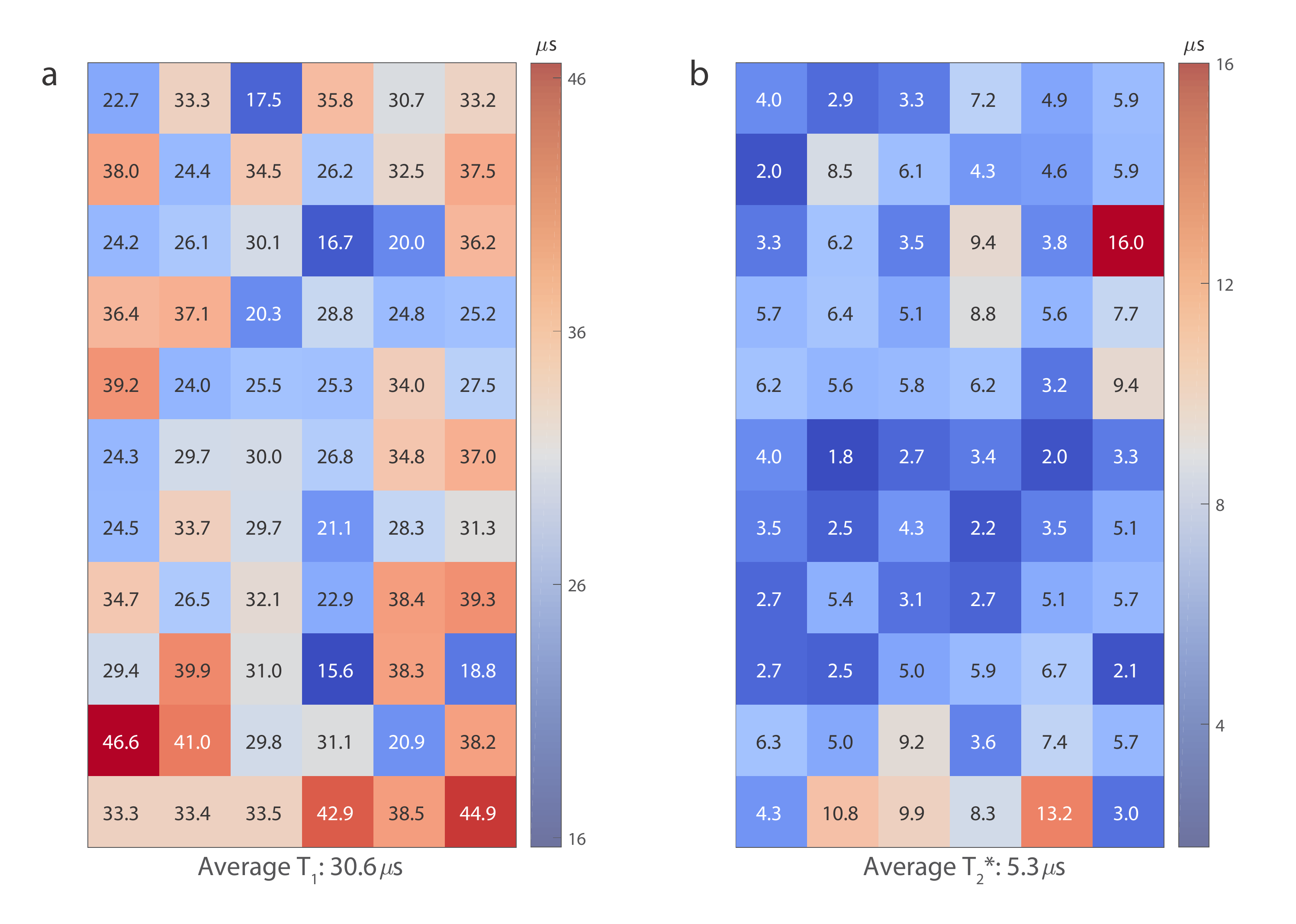}
\end{center}
\caption{\textbf{The relaxation time $T_{1}$ and dephasing time $T_2^*$ of the 66 qubits of the \textit{Zuchongzhi} processor.}
\label{66T1}}
\end{figure*}

\begin{figure*}[!htbp]
\begin{center}
\includegraphics[width=0.65\linewidth]{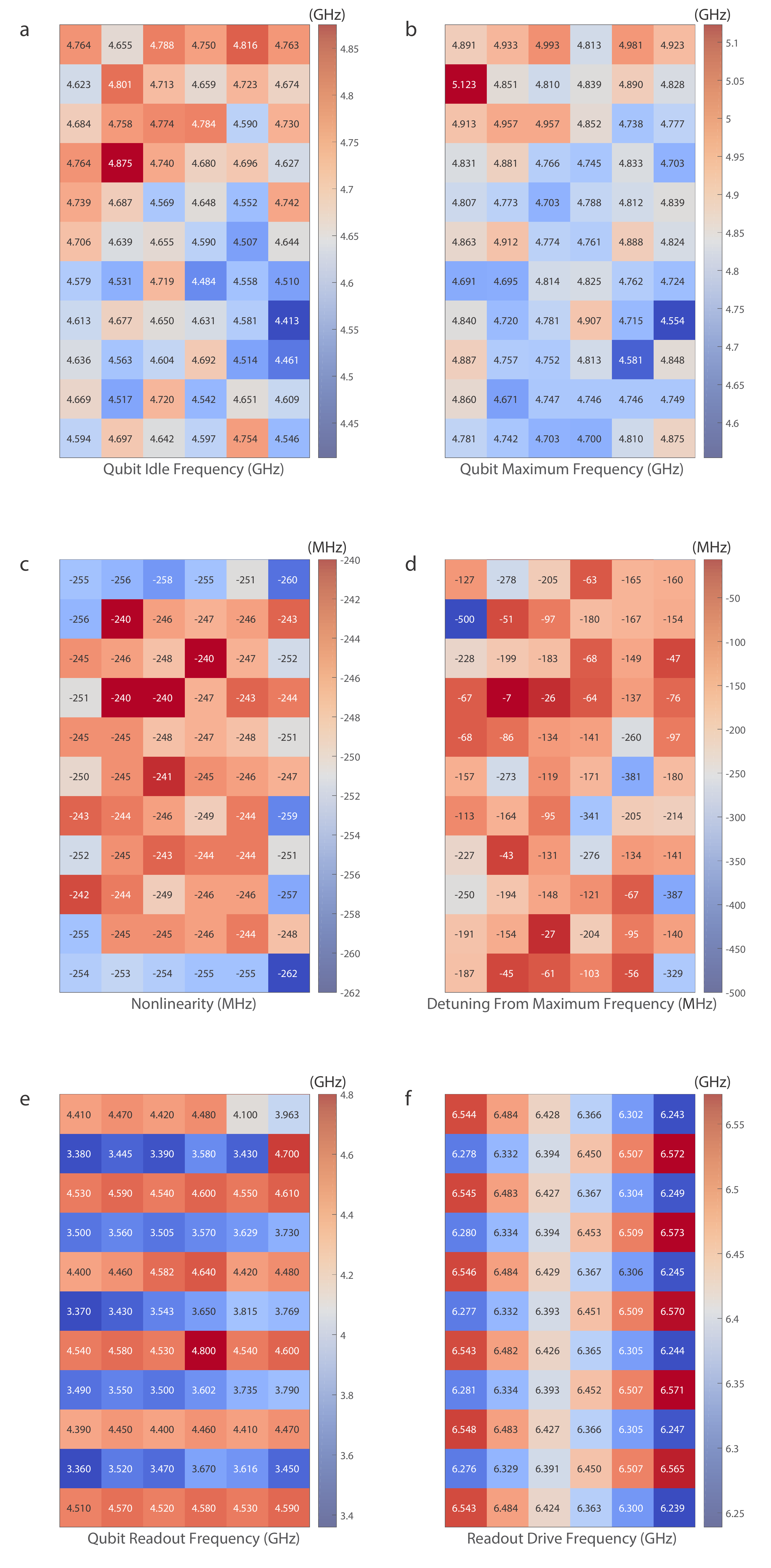}
\end{center}
\caption{\textbf{The typical distribution of 66-qubit parameters over the \textit{Zuchongzhi} processor.}
\label{66qubit}}
\end{figure*}

~\\
\textbf{iii. Two-Qubit Gate Calibration}

The two-qubit gate of our experiment, the iSWAP-like gate~\cite{arute2019quantum}, is realized by tuning the two qubits from their idle frequencies into resonance and turning on the coupling.
%, which can be expressed by the unitary model mention in the main text:
% \begin{equation}
% 	\begin{bmatrix}
% 		1 	& 0 											& 0												& 0 \\
% 		0 	& e^{i(\Delta_{+}+\Delta_{-})}\rm{cos}\theta 		& -ie^{i(\Delta_{+}-\Delta_{-,\text{off}})}\rm{sin}\theta	& 0 \\
% 		0 	& -ie^{i(\Delta_{+}+\Delta_{-,\text{off}})}\rm{sin}\theta	& e^{i(\Delta_{+}+\Delta_{-})}\rm{cos}\theta			& 0 \\
% 		0 	& 0 											& 0												& e^{i(2\Delta_{+}-\phi)}
% 	\end{bmatrix}
% \end{equation}
Considering the effects of decoherence and leakage, we set the total time duration of the iSWAP-like gates to 32ns, which includes the pulse rise/fall time of 3ns and interaction time of 26ns. With this setting, we finely calibrate the detuning of the two qubits and the  coupling strength of the corresponding coupler to obtain a full swap  from one qubit to another. After that, we optimize the interaction time in order to minimize leakage to higher energy levels. We generally perform these calibration steps iteratively until the parameters converge, typically within two or three iterations.

After the calibration of iSwap-like gate paramegters, we benchmark gate performance with XEB. Unlike single qubit gates, iSwap-like gates are parameterized gates each with four parameters as described by the unitary model in the main text, with the experimental results of XEB and the generic unitary model, we learn these parameters through an optimization process to maximize the overall fidelity of XEB.  With the optimized parameters, we fit the final XEB error and speckle purity benchmarking(SPB) error\cite{arute2019quantum}. We obtained high fidelity two-qubit gates with an average XEB pauli error 0.76\% over all 110 couplers when applied simultaneously, which is shown in Fig.~\ref{66error}(c).

\subsection{Part2: Fine-tune on 56 qubits}
\begin{figure}[tbp]
\begin{center}
\includegraphics[width=0.95\linewidth]{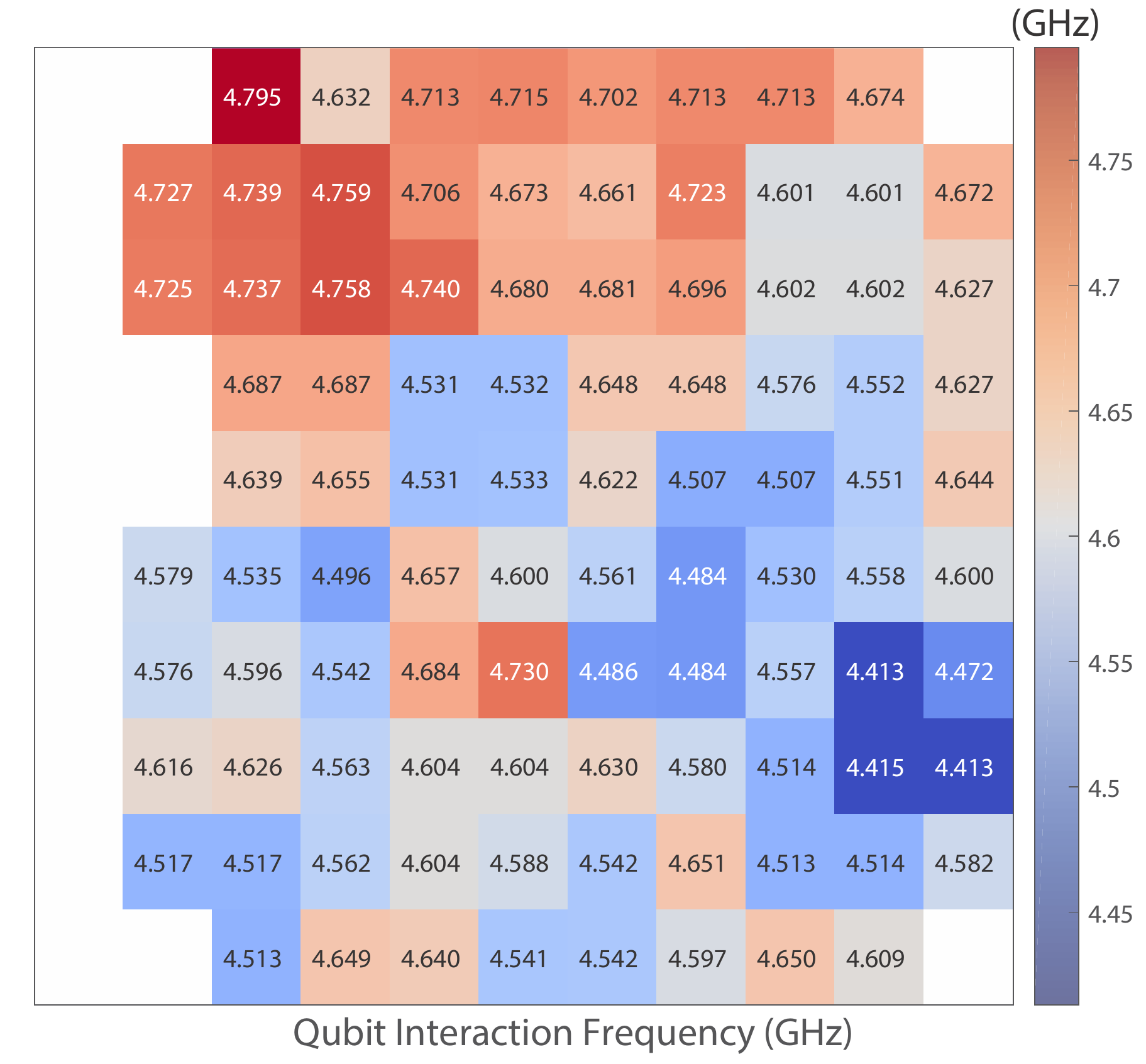}
\end{center}
\caption{\textbf{The qubit interaction frequencies of the 94 two-qubit gate of the 56 selected qubits. }
\label{InteractionFreq}}
\end{figure}

After rough calibrations on 66 qubits, we can estimate the XEB fidelity of a specific quantum circuit with $n$ qubits and $m$ cycles according to Eq.~\ref{fidelitycal}. For random quantum circuit sampling, as the circuit scale increases its fidelity decreases, a large number of samples is required to ensure the uncertainty of the fidelity is much less than the fidelity itself in case of low fidelity. As a trade-off between number of samples and sampling time, we only compile our random circuit sampling tasks to a subset of the processor of up to 56 qubits. We achieved 56-qubit 20-cycle random circuit sampling, an intractable task for classical simulation, within an acceptable sampling time.

To switch from the 66-qubit pattern to the 56-qubit pattern, we just turn off the couplings between the selected qubits and the unused qubits.
Then, we recalibrate the readout, single-qubit gate and two-qubit gate parameters. In addition, for the qubits on which two-qubit XEB pauli error is higher than 1.5\%, we sweep their SPB fidelities at a fixed depth of 20 cycles with 70 random circuit instances at different interaction frequencies to optimize the two-qubit gate fidelities. The best interaction frequencies of the 94 two qubit gates are illustrated in Fig.~\ref{InteractionFreq}. Finally, we use per-layer simultaneous two-qubit gate XEB to benchmark the gate errors~\cite{arute2019quantum}.
%. This step is very  effective, and the final interaction frequency arrangement is shown in Fig.~\ref{InteractionFreq}.

After all the calibrations and optimizations, the average readout error is 4.52\%, the average XEB pauli errors of sinlge-qubit gates and two-qubit gates are 0.14\% and 0.59\%, respectively, as illustrated in Fig.~2 in the main text.

\subsection{Summary of system parameters}
The system parameters of our quantum processor are summarized in Table~\ref{tableSummary}.

\begin{table*}[htb!]
    \centering
    \begin{tabular}{p{6cm}p{2.5cm}<{\centering}p{2.5cm}<{\centering}p{2.5cm}<{\centering}p{2.5cm}<{\centering}}
    \toprule
    Parameters& Median& Mean& Stdev.& Figure\\
    \toprule
  %  \midrule
    Qubit maximum frequency (GHz)& 4.811& 4.809& 0.096& Fig.~\ref{66qubit}\\
    Qubit idle frequency (GHz)& 4.653& 4.653& 0.095& Fig.~\ref{66qubit}\\
    Qubit anharmonicity  (MHz)& -246.5& -248.2& 5.3& Fig.~\ref{66qubit}\\
    Qubit readout frequency (GHz)& 4.400& 4.083& 0.492& Fig.~\ref{66qubit}\\
    Readout drive frequency (GHz)& 6.409& 6.407& 0.103& Fig.~\ref{66qubit}\\
    $T_1$ at idle frequency ($\mu$s)& 30.8& 30.6& 7.1& Fig.~\ref{66T1}\\
    $T_2^*$ at idle frequency ($\mu$s)& 5.1& 5.3& 2.7& Fig.~\ref{66T1}\\
    \hline
    66-qubit readout $e_{r}$ (\%) & 4.52& 4.77& 1.35& Fig.~\ref{66error}\\
    66-qubit 1Q XEB $e_{1}$ (\%) & 0.14& 0.14& 0.05&  Fig.~\ref{66error}\\
    66-qubit 2Q XEB $e_{2}$ (\%) & 0.67& 0.76& 0.43&  Fig.~\ref{66error}\\
    \hline
    56-qubit readout $e_{r}$ (\%) & 4.50& 4.52& 1.43&  Fig.~2(main)\\
    56-qubit 1Q XEB $e_{1}$ (\%) & 0.13& 0.14& 0.05&  Fig.~2(main)\\
    56-qubit 2Q XEB $e_{2}$ (\%) & 0.53& 0.59& 0.20&  Fig.~2(main)\\
   % \bottomrule
    \toprule
    \end{tabular}
    \caption{\textbf{Summary of system parameters.}}
    \label{tableSummary}
    \end{table*}

\section{Software System}
A 66-qubit superconducting quantum processor is a high-dimensional, highly constrained analogy system with system parameters susceptible to environmental changes and often drifts with time, to execute high fidelity quantum circuits, more than 400 DAC, ADC, microwave source and DC source channels need to be controlled with cutting edge precision. Operating such a complex quantum system requires considerable advancement in software as compared to  operating small quantum systems. We have developed a software system for intermediate scale superconducting quantum systems called \textit{QOS(Quantum Operating System)} that is capable of operating quantum systems of more than 1000 qubits.

\begin{figure}[!htbp]
\begin{center}
\includegraphics[width=0.55\linewidth]{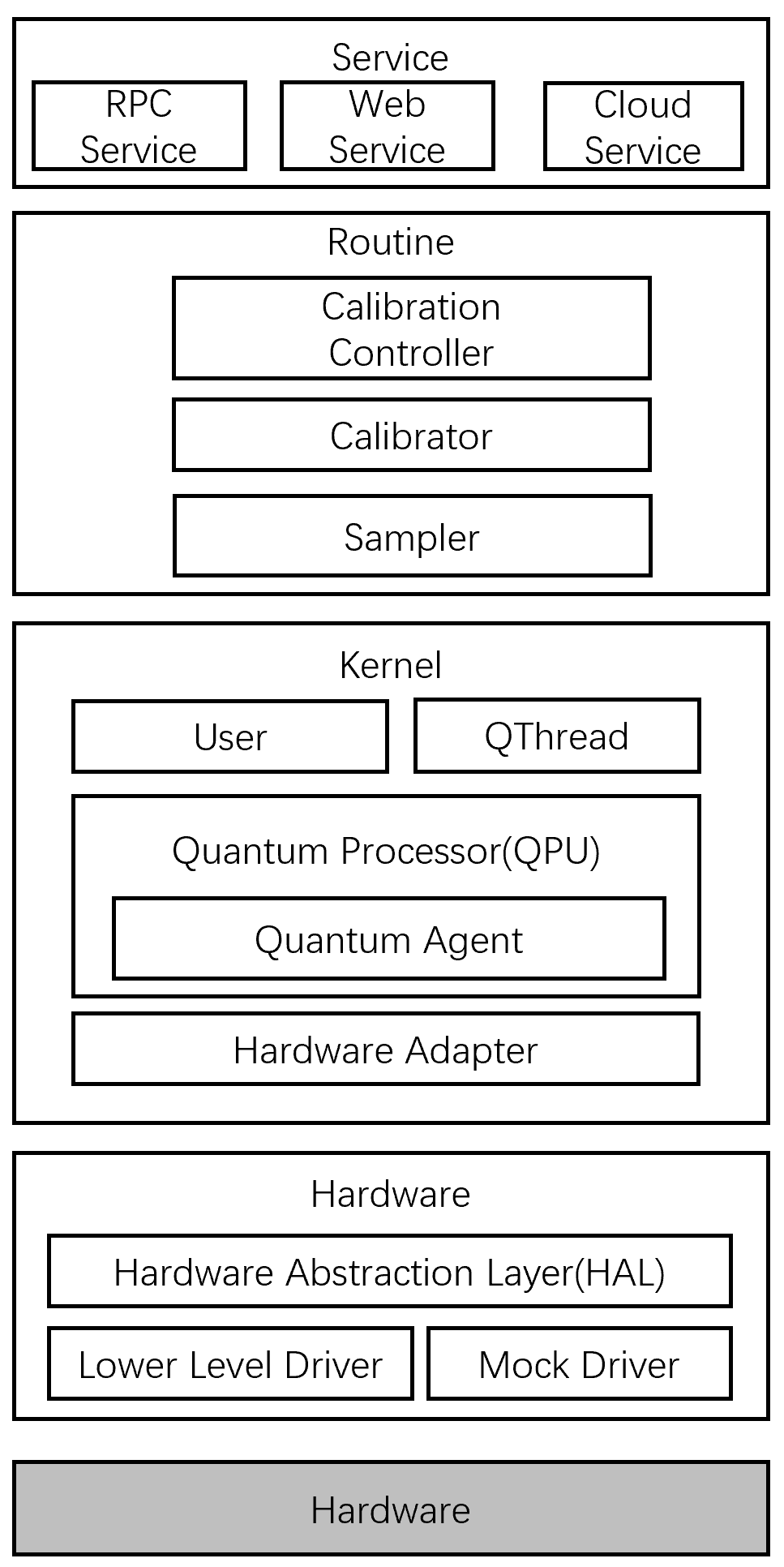}
\end{center}
\caption{\textbf{Software architecture.}  The \textit{QOS} system consists of four major modules, the \textit{Hardware} module , the \textit{Kernel} module, the \textit{Routine} module and the \textit{Service} module. Other supporting modules are not shown.
\label{software_architecture}}
\end{figure}

The major functionalities of \textit{QOS} are abstract away hardware details, manage resources and implement quantum operations. The first two functionalities are similar to that of a classical computer operating system, the last is unique to a quantum operating system.

As mentioned before, a large scale quantum processor is a complicated system, directly managing such a complex system is intractable for experiment users, the \textit{QOS} system abstract away system details and manages all the  resources for the user. Resources of a quantum computer system includes hardware and software. We categorize hardware into two categories, \textit{classical hardware} and \textit{quantum hardware}. \textit{Classical hardware} includes DAC, ADC, microwave source, DC source and all other control electronics, \textit{quantum hardware} includes the quantum processor and the quantum amplifiers. To mange all the resources, \textit{QOS} manages a \textit{registry} system to represent and keep track of the settings/configurations of all the resources.  The \textit{registry} is organized in a tree structure and holds all the settings of the \textit{classical hardware}, \textit{quantum hardware} and software configurations. Each user maintains independent \textit{registry} settings for their own experiments. About 20k entries of \textit{registry} settings are used to run the quantum processor in this work.

Fig.~\ref{software_architecture} shows the basic architecture of the the QOS system, it consists of four major modules, the \textit{Hardware} module, the \textit{Kernel} module, the \textit{Routine} module and  the \textit{Service} module. Other supporting modules includes the \textit{Data Storage} module, the \textit{Registry} module, the \textit{Waveform} module and the \textit{Utility} module etc. Most of the software components are  abstractions of the hardware system components.

The \textit{Hardware} module hosts a library of drivers of classical hardware, on top of this, a hardware abstraction layer(\textit{HAL}) is introduced to provide an unified API for the \textit{Kernel} module to interface with hardware, making upper stream modules hardware model agnostic and can be adapted to hardware provided by differ vendors with minimum effort. Mock drivers are also provided to facilitate \textit{off-line} testing without real hardware.

Unlike a classical processor, on which all the basic computing units like logical gates and registers are hardware components, a superconducting quantum processor is just a physical quantum system, all computing units like quantum gates and quantum registers have to be implemented by software. The most essential functionality of a quantum software system is to implement quantum operations and transform a quantum physical device into a quantum computing processor. This is done by the \textit{Kernel} module and makes it the most important module. The \textit{Kernel} module is an abstraction of the physical system, including the quantum processor device and the controlling electronics, it consists of the following sub-modules, the \textit{Quantum Processor(QPU)}, the \textit{Hardware Adapter}, the \textit{User} and the \textit{QThread}.

Controlling superconducting qubits to implement high fidelity quantum gates is a complex task, when the system size scales up, the requirement on software performance is very demanding. To be scalable, the software system has to be able parse quantum operations in parallel. Entanglement is an intrinsic feature of a quantum system, implement parallel operation on an entangled system is a non trivial task. We use the \textit{Actor Model}~\cite{hewitt1973universal} for the implementation of the \textit{kernel} module. Each hardware component is modeled as an actor, called \textit{Agent} with those representing quantum components called \textit{QAgent} in particular, types of \textit{agent} used in this work are listed below:

\begin{itemize}

  \item \textit{Transmon}, of the \textit{QAgent} class, each represents a Transmon~\cite{koch:042319} type or Transmon type compatible qubit, responsible for managing all the settings of the qubit and implementing single qubit quantum operations.

  \item \textit{TTG}, tunable coupler of the Transmon qubit as used on the \textit{Zuchongzhi} processor, responsible for managing all the settings of the coupler and implementing two qubit quantum operations.

  \item \textit{MuxReadout}, represents a multiplexed qubit state readout unit.

  \item \textit{IMPA}, represents an impedance matched parameter amplifier~\cite{mutus2014strong}.

  \item \textit{QProcessor}, a controlling \textit{agent} of all the agents, including qubits, couplers, readouts, amplifiers, hardware manager etc.

  \item \textit{Hardware Adapter}, represents one piece of classical hardware, DAC, ADC, microwave source etc.

  \item \textit{Hardware Manager}, a controlling \textit{agent} of all the classical hardware agents.

  \item \textit{QThread}, a software abstraction that represents a quantum thread or quantum process on a quantum processor.

  \item \textit{User}, a software abstraction that is responsible for managing user specific resources, conducting user transactions like authentication and authorization etc.

\end{itemize}

\begin{table*}[!htbp]
    % \centering
    % \begin{tabular}{l{5em}|l{5em}|l{5em}}
    \begin{tabular}{l|l|l}
    \toprule
    \textbf{Opcode} & \textbf{Example} & \textbf{Meaning}\\
    % \toprule
    % \midrule
    \hline
    SET & SET Q01 2 3.14 & Set value of the selected setting entry of index 2 of Q01 to 3.14\\
    \hline
    CMT & CMT Q01 0 & Commit settings change, waveforms regenerated, caches refreshed etc. \\
     & & Example: commit settings change of type 0 for Q01\\
    \hline
    I & I Q01 50 & Idle gate. Example: Q01 idle for time duration 50\\
    \hline
    X, Y, Z & X2M Q01 & XY gates. Example: -X/2 gate on Q01\\
    X2P, X2M, Y2P, Y2M & &\\
    \hline
    H & H Q01 & H gate\\
    \hline
    XY & XY Q01 0.785 & $\pi$ rotation around the axis of azimuth 0.785 rad in the XY plane\\
    \hline
    XY2P, XY2M & XY2P Q01 0.785 & $+\pi/2$($-\pi/2$) rotation around the axis of azimuth 0.785 rad\\
    & XY2M Q01 0.785 & in the XY plane\\
    \hline
    X12 & X12 Q01 & $|1\rangle \leftrightarrow |2\rangle$ driving gate \\
    \hline
    X23 & X23 Q01 & $|2\rangle \leftrightarrow |3\rangle$ driving gate \\
    \hline
    Z & Z Q01 & Z gate on Q01\\
    \hline
    S, SD & SD Q01 & $S$, $S^{\dagger}$ gate \\
    \hline
    T, TD & T Q01 & $T$, $T^{\dagger}$ gate\\
    \hline
    RX, RY, RZ & RY Q01 0.785 & Arbitrary rotation around the X, Y or Z axis with the specified angle\\
    \hline
    RXY & RXY Q01 0.785 3.14 & Q01 rotation of 3.14 rad around the the axis of azimuth angle 0.785 \\
    & & in the XY-plane \\
    \hline
    AXY & AXY Q01 20 0.75 0.785 -1e6 0.55 & Arbitary XY rotation. Example: Q01 arbitrary rotation around the the axis of \\
    & & azimuth angle 0.785 in the XY-plane with pulse length 20,\\
    & & pulse amplitude 0.75, -1e6 detuning and 0.55 DRAG alpha\\
    \hline
    CZ, CNOT, SWP, ISWP& CZ G0701 & Two qubit gates. Example: CZ gate on coupler G0701 \\
    SISWP & & \\
    \hline
    CP, FSIM & FSIM G0701 1 & Parameterized two qubit gates, with an index to indicate which \\
     & & parameter set to use. Example: FSIM gate of index 1 on coupler G0701 \\
    \hline
    DTN & DTN Q01 100 -2e6 0 & Detune Q01 -2e6MHz for a duration 100 at time offset 0\\
    \hline
    MEASURE, M & M Q01 & Measure qubit Q01 \\
    \hline
    PLS, PLSXY & PLS Q01 1 100 10 0.8 0 0 1 & Put a pulse of a waveform index 1 at time 100 with the specified\\
     & & parameters for Q01 \\
    \hline
    B & B Q01 R01 & Establish a time barrier between Q01 and R01\\
    \hline
    SWD, SWA & SWA Q01 0.8 & SET work bias duration, amplitude. The work bias amplitude \\
    & & determines $f_{01}$\\
    \hline
    MOV,~... & MOV G0701 0 1 & Classical instructions. Example: set value 1 to classical register 0 of G0701\\
    \toprule
    \end{tabular}
    \caption{\textbf{Quantum Control Instruction Set (QCIS)}}
    \label{qcis}
\end{table*}

By introducing the concept of \textit{agent}, we decouple the highly entangle quantum processor as well as the shared classical controlling electronics into a collections of separated components, with all the entanglement resolved by the related agents and transparent to upper stream modules. We use a highly decentralized architecture for the agents, all agents function concurrently and directly communicate with each other when necessary. For the only tow central agents, the \textit{QProcessor} and the \textit{Hardware Manager}, they only assume minimum light weighted functionalities like instruction dispatching and synchronization to avoid any performance bottle neck. With this architecture, we can process large scale complex quantum operations in a fully concurrent manner, pushing software performance to its limits.

Multi-processing or multi-threading is supported through \textit{QThread}, one can run multiple experiments concurrently, new experiments can be created and executed immediately without waiting for previous experiments to finished. When an experiment is started, typically one \textit{QThread} is created with a new \textit{context} holding a view of the \textit{registry} and other dynamic settings in each involved \textit{QAgent}, exposing only a few selected setting entries to be dynamically set for parameter sweeping. Tasks of this experiment is configured with settings from these contexts and executed through time slicing with other \textit{QThread}s with round-robin scheduling. Several \textit{QThread}s operating upon distinct \textit{agent} sets can be configured to run simultaneously to save hardware time. A \textit{QThread} will go into hibernation after idling for some time to yield resources, and will be awakened upon arrival of new tasks. Multi-user is also support through \textit{user}, multiple users can run experiments concurrently with completely different settings. \text{QOS} works in a server-client mode, a dedicated RPC(Remote Procedure Call) framework is developed for users to interface with \text{QOS}, web service and cloud service are also supported.

We introduce the \textit{Quantum Control Instruction Set(QCIS)} for the full control of a superconducting quantum processor as list in table~\ref{qcis}. \textit{QCIS} offers a unified API for operating a quantum processor, with \textit{QCIS}, we can control the \textit{Zuchongzhi} system exactly the same way as programming a classical computer. Program written with \textit{QCIS} are called \textit{QProgram}. All experiments in this work are implemented with this instruction set. As an example, sweep with the following code snippet executes a tuning of all the iSWAP-like gate related parameters for a swap experiment on \textit{G0701}, the \textit{TTG QAgent} between qubits \textit{Q01} and \textit{Q07}:
\lstset{ % General setup for the package
    language=Python,
    % backgroundcolor=\color{yellow!20},
    morekeywords={SET, X2P, Y2P, XY2P, MOV, CMT, X, FSIM, B, M},
    basicstyle=\tiny\sffamily,
    % numbers=left,
     numberstyle=\tiny,
    frame=tb,
    tabsize=4,
    columns=fixed,
    showstringspaces=false,
    showtabs=false,
    keepspaces,
    commentstyle=\color{ForestGreen},
    comment=[l]{//},
    morecomment=[s]{/*}{*/},
    commentstyle=\color{purple}\ttfamily,
  % stringstyle=\color{red}\ttfamily,
    keywordstyle=\color{blue}\bfseries
}
\begin{lstlisting}[
    basicstyle=\tiny, %or \tiny \small or \footnotesize etc.
]
SET G0701 0 30      // set FSIM gate duration to 30*
SET G0701 1 6       // set qubit pulse rise/fall edge duration to 6
SET G0701 2 2       // set coupler pulse rise/fall edge duration to 2
SET G0701 3 -10e6   // set FSIM gate qubit_0 detune to -10e6
SET G0701 4 20e6    // set FSIM gate qubit_1 detune to 20e6
SET G0701 5 15e6    // set coupler coupling strength to 15e6
MOV G0701 0 1       // set value 1 to classical register 0 to indicate that the
                    // index of  the FSIM gate for the following CMT action is 1
CMT G0701 0         // commit settings change, waveforms will be regenerated,
                    // caches refreshed etc.
X Q07               // X gate on Q07
FSIM G0701 1        // execute the FSIM gate indexed 1 on G0701**
B Q07 Q01           // establish a time barrier before measurement
B Q07 R01           // establish a time barrier before measurement
M Q07               // measure Q07
M Q01               // measure Q01
/*
*: at the creation of QThread, related setting entries are selected and indexed
**: a coupler can implement multiple FSIM gates of different parameters
*/
\end{lstlisting}

The following \textit{QProgram} executes one of the random circuit sampling task in this work:

\begin{lstlisting}[
    basicstyle=\tiny, %or \tiny \small or \footnotesize etc.
]
X2P Q01
Y2P Q02
X2P Q03
Y2P Q04
XY2P Q07 0.785398163397448
X2P Q08
Y2P Q09
XY2P Q10 0.785398163397448
X2P Q11
XY2P Q12 0.785398163397448
XY2P Q13 0.785398163397448
X2P Q14
XY2P Q15 0.785398163397448
X2P Q16
XY2P Q17 0.785398163397448
XY2P Q19 0.785398163397448
X2P Q20
XY2P Q21 0.785398163397448
XY2P Q22 0.785398163397448
Y2P Q23
XY2P Q25 0.785398163397448
XY2P Q26 0.785398163397448
// ... 30 lines not shown
XY2P Q61 0.785398163397448
XY2P Q62 0.785398163397448
Y2P Q63
Y2P Q64
FSIM G1003 1
FSIM G1104 1
FSIM G1408 1
FSIM G1509 1
// ... 1809 lines not shown
M Q50
M Q51
M Q52
M Q53
M Q55
M Q56
M Q57
M Q58
M Q59
M Q61
M Q62
M Q63
M Q64
\end{lstlisting}

The \textit{Routine} module is a standard library of quantum experiments, implemented mainly for bringing up a superconducting quantum processor. Basic routines are called \textit{Sampler}s. A \textit{sampler} creates a \textit{QThread}, provides a task generator for the \textit{QThread} which then throttles tasks to \textit{QPU} to be executed. Measured results are send back as an asynchronous stream, which can be subscribed by data handlers to do data processing on the fly. Data handlers could be simple data saving routines, data visualizers or sophisticated pattern recognition routines in data analyzers of a calibrator. All \textit{sampler}s except a few special cases are implemented to run experiments for multiple \textit{QAgent}s in parallel. Upon \textit{sampler}s, \textit{calibrator}s are build to implement a specific calibration task, a \textit{calibrator} builds one or several \textit{sampler}s with parameters like sweep ranges defined by the specific calibration task, run the sampling tasks, analyze the sampled data and generate a calibration result \textit{CalResult} for each of its calibration target. The calibration results are then validate by one or several \textit{Validator}s to check whether they are solid. On top of \textit{calibrator}s, a \textit{Calibration Controller} manages all the \textit{calibrator}s and schedules then to run in a specific scheme, collects calibration results from different calibrators, render results into \textit{Calibration Report} and update related setting entries according to the results. The \textit{calibration controller} can run in several modes, fully automated run as a directed graph~\cite{kelly2018physical}, run selected \textit{calibrators} in a specified series or just run a single \textit{calibrator}. High flexibly is provided in calibration mode to meet the needs of  experiments, a \textit{calibrator} can be triggered to run a partial calibration on a subset of its responsible \textit{agents} or run a series of calibration dividing then into subgroups as opposed to calibrate all \textit{agents} in parallel. With \textit{calibrator}s and \textit{calibration controller}, the bring up/calibration procedure can be largely automated.

\begin{figure}[!htbp]
\begin{center}
\includegraphics[width=0.95\linewidth]{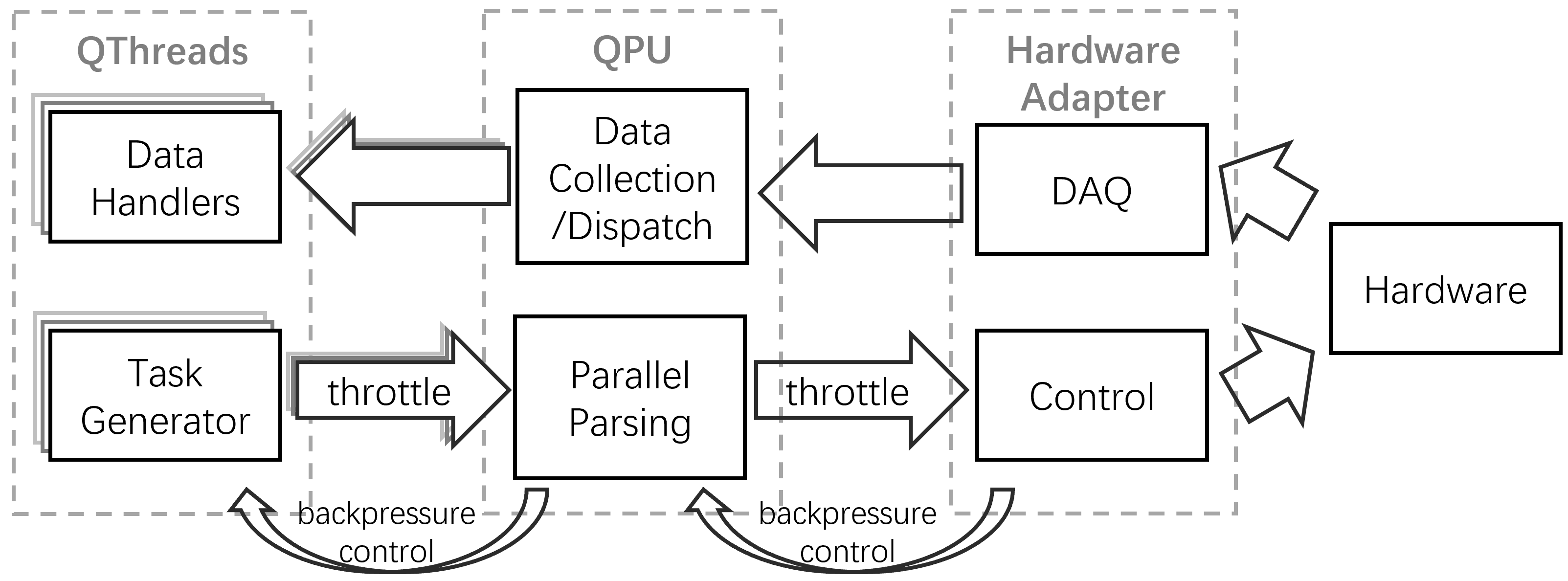}
\end{center}
\caption{\textbf{Experiment workflow.} The \textit{QOS} system works in a pipe-lined mode with back pressure control, together with parallel processing at all pipeline stages, software overhead is greatly reduced.
\label{fig_s_software_2}}
\end{figure}

As a superconducting quantum system is unstable and its parameters drifts with time, fast calibration is key to success. Performance is not just crucial for the \textit{kernel}, we also designed all the \textit{sampler}s and \textit{calibrator}s with high performance in mind. We use a pipe-lined workflow for the whole system as show in FIG.~\ref{fig_s_software_2} to minimize software processing overhead. As a result, \textit{QOS} can run heavy weight experiments like parallel randomized benchmarking on large scale quantum processor with negligible software time overhead. In this work of 66 qubits and 110 couplers, we can perform one round of routine calibration in less than one hour, which includes single qubit gate calibration, single qubit gate XEB fidelity calibration for all the qubits, two qubit gate calibration, two qubit gate XEB fidelity calibration and learn iSWAP-like gate parameters for all the couplers. This experiment time consumption will not grow considerably as it is only limited by hardware time.

\section{Random quantum circuits}

In each cycle of the random quantum circuit, single-qubit gates chosen randomly from $\{\sqrt X, \sqrt Y, \sqrt W \}$ are first applied on all qubits, and then two-qubit gates, iSWAP-like gate, are applied to pairs of qubits. There are four types of patterns for two-qubit gate, labeled by A, B, C and D respectively, which determine which two-qubit gates are executed on each layer. Different from the single-qubit gates, the four patterns of A, B, C, and D are implemented in the sequence of ABCDCDAB.
%Executing these four patterns sequentially from A to D once is equivalent to executing all the two-qubit gates once.
%The structure of these four patterns could affect the hardness of simulating the RQC on classical computer, and thus should be carefully designed.

\begin{figure}[!htbp]
\begin{center}
\includegraphics[width=0.6\linewidth]{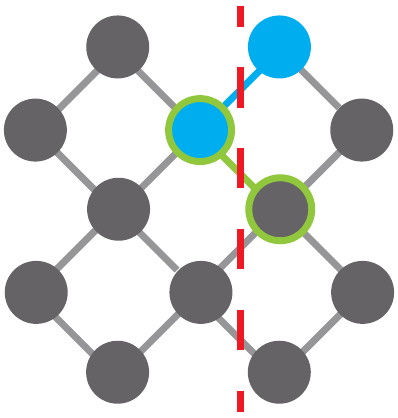}
\end{center}
\caption{\textbf{Wedge formation.} The gate highlighted in blue and green on the cut line are crosspartition gates. A wedge is formed when two consecutive crosspartition gates share a qubit.
\label{wedge}}
\end{figure}

In our experiments, we implement random quantum circuits (RQCs) on the quantum processor, and evaluate the time consumption of simulating these quantum circuits on the traditional supercomputer. So far, there are mainly two types of algorithms for simulating large RQCs: (1) tensor network algorithms~\cite{MarkovShi2008,GuoWu2019,VillalongaMandra2019,villalonga2020establishing,HuangChen2020,pan2021simulating,guo2021verifying} and (2) Schr{\"o}dinger-Feynman algorithm (SFA). The time-consuming of the SFA is significantly connected to the structure of four patterns A, B, C, and D, which could be regarded as a criteria for designing the structure of RQCs. The basic procedure of SFA can be summarized as follows: (1) Firstly, cut a $n$-qubit quantum circuit into two partitions with $n_1$ and $n_2$ qubits. (2) Then by summing all simulation paths that are the product of the terms of the Schmidt decomposition of all cross-partition gates, one could obtain the output state of the quantum circuit. The computational complexity of the algorithm is proportional to $(2^{n_1} + 2^{n_2})r^g$, where $r$ is the Schmidt rank, and $g$ is the number of the cross-partition gates. The iSWAP-like gate used in our scheme has the Schmidt rank of $r=4$. Thus, when evaluating the runtime of the SFA, one needs to find the optimal cut with $n_1$, $n_2$ and $g$ that makes the simulation task to become the easiest. In addition, for a circuit cut, if the following formations occur, both $r$ and $g$ may be reduced, resulting in a reduction in the computational complexity of the SFA algorithm:

1) \textit{Wedge formation}. As shown in Fig.~\ref{wedge}, a wedge formation is formed when two consecutive cross-partition gates share a qubit, which can reduce the Schmidt decomposition of resulting three-qubit unitary to only four terms. Equivalently, every wedge reduces a cross-partition gate, and provides a speedup of a factor of 4.

2) \textit{DCD formation}. The DCD formation often happens at the boundary. Specifically, the DCD formation appears when there are three successive two-qubit gates acting on the qubit pairs $(a,b)$, $(b,c)$ and $(a,b)$, and these three gates can be fused in one (the two gates on qubit pairs $(a,b)$ can be fused). A DCD formation provides a speedup of a factor of 4.

3) \textit{Formation of iSWAP-like gates at the start and end of the circuit}. The iSWAP-like gate is the product of a iSWAP and controlled phase gate, and the iSWAP gate can be applied either at the beginning or at the end of the sequence. We apply this transformation to all iSWAP-like gates at the beginning (end) of the circuit that affect qubits that are not affected by any other two-qubit gate before (after) in the circuit. The iSWAP is then applied to the input (output) qubits and their respective one-qubit gates trivially, and the bond dimension remaining from this iSWAP-like gate is 2, corresponding to a controlled phase gate, as opposed to the bond dimension 4 of the original iSWAP-like gate. Thus, A iSWAP-like gate that appears at the same time in the cross-partition and the beginning (or end) of the circuit can provide a speedup of a factor of 2.

%\begin{figure}[htbp]
%\begin{center}
%\includegraphics[width=0.9\linewidth]{undirected_graph}
%\end{center}
%\caption{\textbf{The undirected graphical model of random quantum circuit} (a) Random quantum circuit (b) The undirected graphical model of random quantum circuit. Each non-diagonal single-qubit gate introduces a new vertex or variable. Note that, even though two-qubit gates are generally represented by a clique with four vertices or variables, Sycamore gates can be simplified as a cphase followed by a SWAP. The cphase is represented as an edge between two existing variables. The SWAP, however, provides more complexity to the graph as it swaps the corresponding variables.
%\label{undirectedgraph}}
%\end{figure}

To ensure the computational complexity and classical hardness with low depth, the number of cross-partition gates $g$ on the optimal cut should be large enough, and the four patterns should be carefully designed to avoid the occurrence of above three formations as much as possible.

\begin{figure}[!htbp]
\begin{center}
\includegraphics[width=1\linewidth]{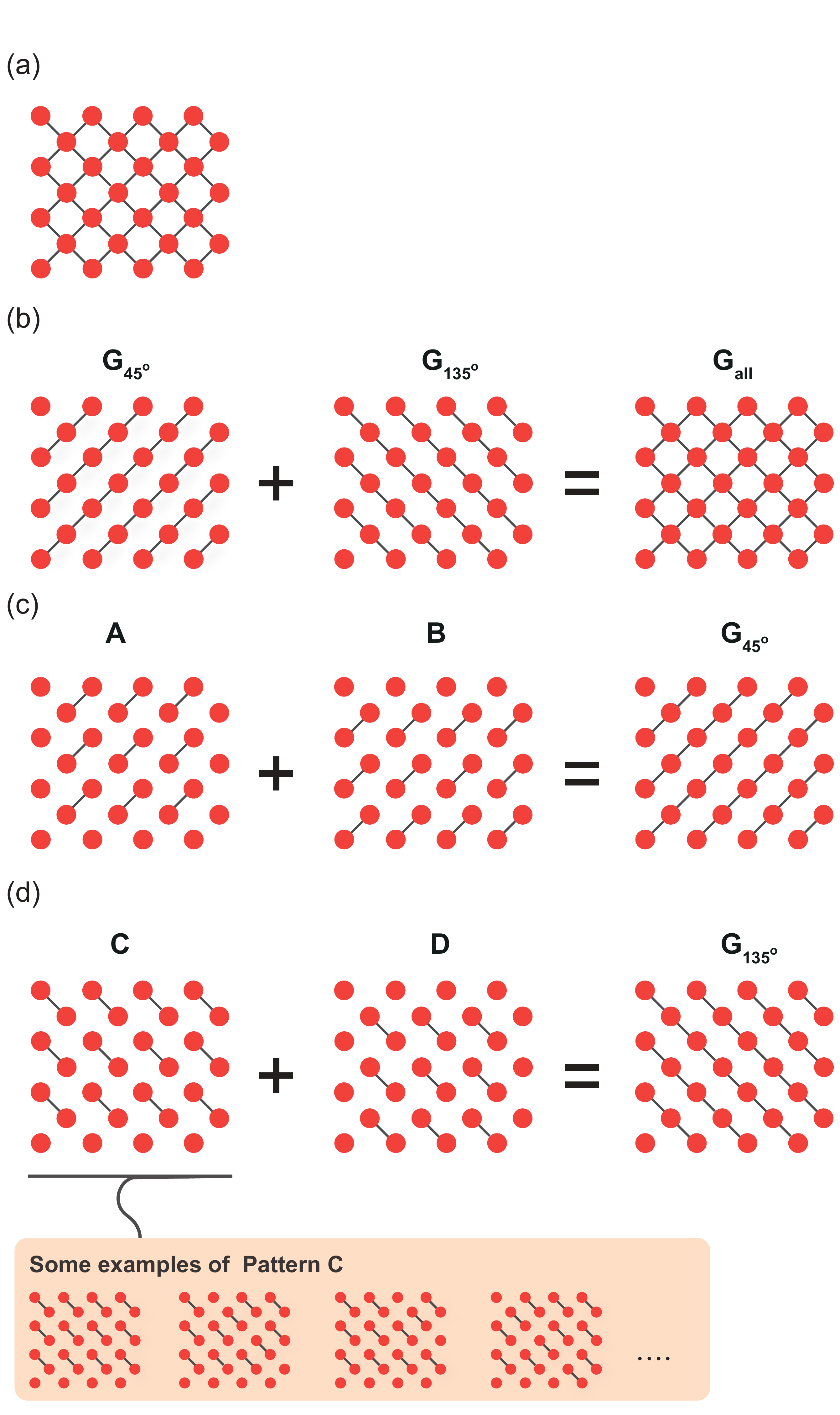}
\end{center}
\caption{\textbf{The relationship of A, B, C, and D patterns.} (a) An example of the topology of a quantum processor. The red dot represents the qubit, and the black line represents the two-qubit gate between two qubits. (b) These two-qubit gates can be divided into two categories: ${G_{45^\circ }}$ and ${G_{135^\circ }}$, which are represented as 45 degree lines and 135 degree lines, respectively. (c) The pattern ${G_{45^\circ }}$ is split into pattern A and pattern B. (d) The pattern ${G_{135^\circ }}$ is split into pattern C and pattern D. Patterns A, B, C, and D could have different structures, here we list some examples of pattern C.
\label{ABCD}}
\end{figure}

Figure~\ref{ABCD}(a) shows an example of the topology of a quantum processor. The red dot represents the qubit, and the black line represents the two-qubit gate between two qubits. These two-qubit gates can be divided into two categories: ${G_{45^\circ }}$ and ${G_{135^\circ }}$, which are represented as 45 degree lines (see Fig.~\ref{ABCD}(b)) and 135 degree lines (see Fig.~\ref{ABCD}(b)), respectively. For efficient implementation on the quantum hardware, every two different two-qubit gates in each pattern should not share the same qubit. Thus, pattern ${G_{45^\circ }}$ is split into pattern A and pattern B (see Fig.~\ref{ABCD}(c)). Similarly, pattern ${G_{135^\circ }}$ is split into pattern C and pattern D (see Fig.~\ref{ABCD}(d)). Patterns A, B, C, and D could have different structures, and once pattern A (or C) is determined, then the structure of pattern B (or D) is determined. To design the optimal patterns of A, B, C, and D for a specific quantum processor, a search strategy is proposed as shown below:

(1) Set constraints, including the topology of the processor, the circuit depth of RQC, the sequence of four patterns (usually ABCDCDAB), and the maximum permissible number of two-qubit gates in the partition.

(2) Set the structures of patterns of A, B, C, and D.

(3) According to the conditions in (1) and (2), search for the optimal cuts with the least number of effective crosspartition gates, where the number of effective crosspartition gates $L$ is determined using the following formula,
\begin{align}
L=g_{\text{cut}}-g_{\text{wedge}}-g_{\text{DCD}}-\frac{g_{{\text{start,end}}}}{2}
\label{eq:1}
\end{align}
where $g_{\text{cut}}$ is the number of crosspartition gates, $g_{\text{wedge}}$ is the number of wedge formations, $g_{\text{DCD}}$ is the number of DCD formations, and $g_{{\text{start,end}}}$ is the number of the formations of iSWAP-like gates at the start and end of the circuit. Step (3) outputs the optimal cut and the number of corresponding effective crosspartition gates, denoted as $\text{Min}_L$, which determines the computational complexity of SFA for simulating the circuit designed with the set patterns.

\begin{figure*}[htbp]
\begin{center}
\includegraphics[width=0.95\linewidth]{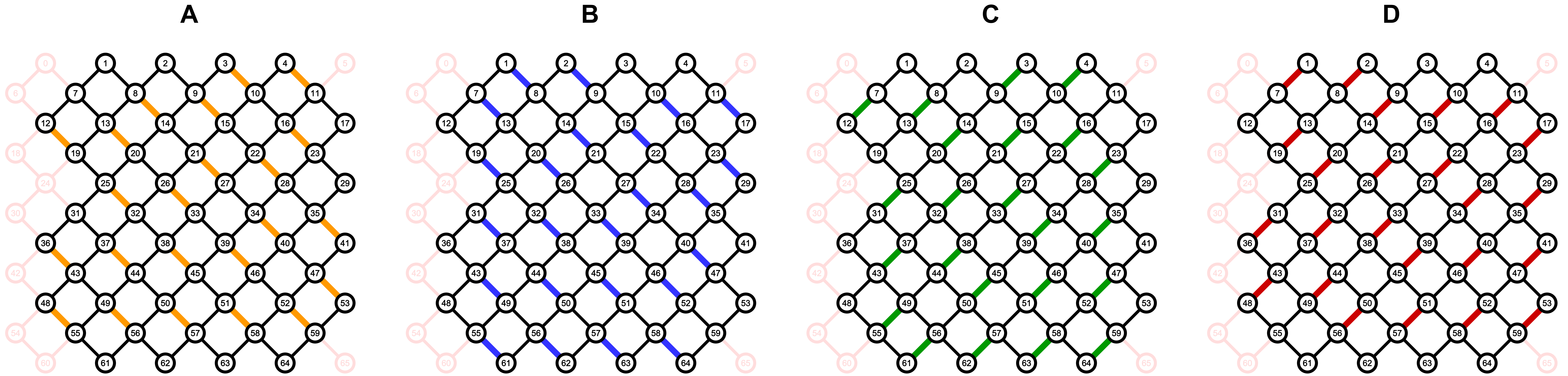}
\end{center}
\caption{\textbf{Coupler activation patterns for 56 qubits.} Coupler activation pattern for 56 qubits that determines which qubits are allowed to interact simultaneously in a cycle.
\label{66circuit}}
\end{figure*}

\begin{figure}[htbp]
\begin{center}
\includegraphics[width=0.85\linewidth]{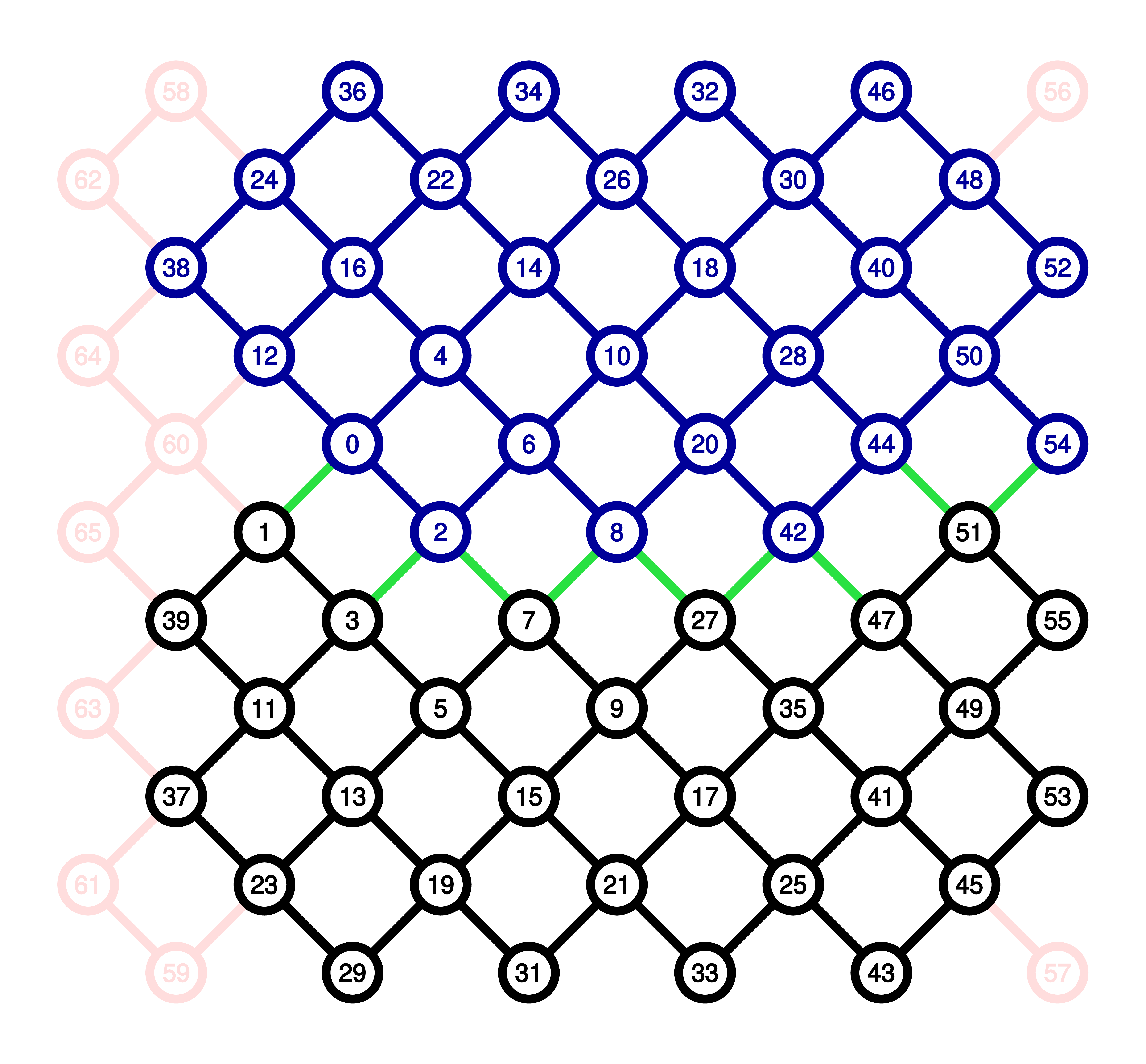}
\end{center}
\caption{\textbf{Qubit ordering and optimal cut for 56-qubit circuit with 20 cycles.} This order determines which qubits are used for $n$ qubits experiments. The green lines form the optimal cut for 56-qubit circuit with 20 cycles.
\label{Qubit ordering}}
\end{figure}

%We will design different sequences for a circuit depth.
Repeat steps (2) and (3) to search for the optimal patterns that have the  maximum $\text{Min}_L$. Figure~\ref{66circuit} and Figure~\ref{Qubit ordering} show the optimal patterns of A, B, C, and D that we have searched for 56-qubit RQC with 20 cycles and the corresponding promising cut, respectively.
%Finally, We select the optimal sequence and the structures of patterns of A, B, C, and D for given constraints.
%We note that for the last two patterns of quantum circuits with depths of 14 and 18, we no longer follow the sequence of ABCDCDAB. Specifically, for the depths of 14 and 18, the sequence is selected as ABCDCDABABCDCA and ABCDCDABABCDCDBABD respectively, to ensure more effective crosspartition gates at promising cut.

\section{XEB result analysis}
\subsection{XEB fidelity}
%In theory, circuits of enough depth exhibit the Porter-Thomas distribution.
For a set of bitstrings $\{q_i\}$, the cross-entropy benchmarking (XEB) fidelity is estimated from the ideal probabilities $\{p_i=p_s(q_i)\}$ as
\begin{equation}
F_l = \langle Dp\rangle  - 1
\label{linearXEB}
\end{equation}
\begin{equation}
F_c = \langle \log (Dp)\rangle  + \gamma
\label{logarithmicXEB}
\end{equation}
where $F_l$ is the linear XEB, $F_c$ is logarithmic XEB, and $\gamma\approx 0.577$ is the Euler-Mascheroni constant.

\subsection{Prediction of circuit fidelity}
As shown in Ref.~\cite{arute2019quantum}, the predicted fidelity $F$ could be calculated from a simple multiplication of individual operation fidelities as

\begin{equation}
F{\rm{ = }}\prod\limits_{g \in {G_1}} {(1 - {e_g})} \prod\limits_{g \in {G_2}} {(1 - {e_g})} \prod\limits_{q \in Q} {(1 - {e_q})}
\label{fidelitycal}
\end{equation}
where $e_g$ are the individual gate Pauli errors, $G_1$ and $G_2$ are the set of single-qubit gates and two-qubit gates, respectively, $Q$ is the set of qubits, and $e_q$ are the state preparation and measurement errors of individual qubits.

\subsection{Performance of patch circuits and elided circuits}

Considering that it is unrealistic to verify the fidelity of the large-scale quantum circuits on current supercomputer, thus two verifiable quantum circuits, patch circuits and elided circuits, are introduced to provide approximate predictions of system performance. In the patch circuits, all two-qubit gates across the partitions are elided, which is essentially two disconnected circuits running in parallel. Compared to patch circuits, only a fraction of the two-qubit gates along the cut during a few early cycles of the sequence is elided in the elided circuits.

\begin{figure}[htbp]
\begin{center}
\includegraphics[width=0.9\linewidth]{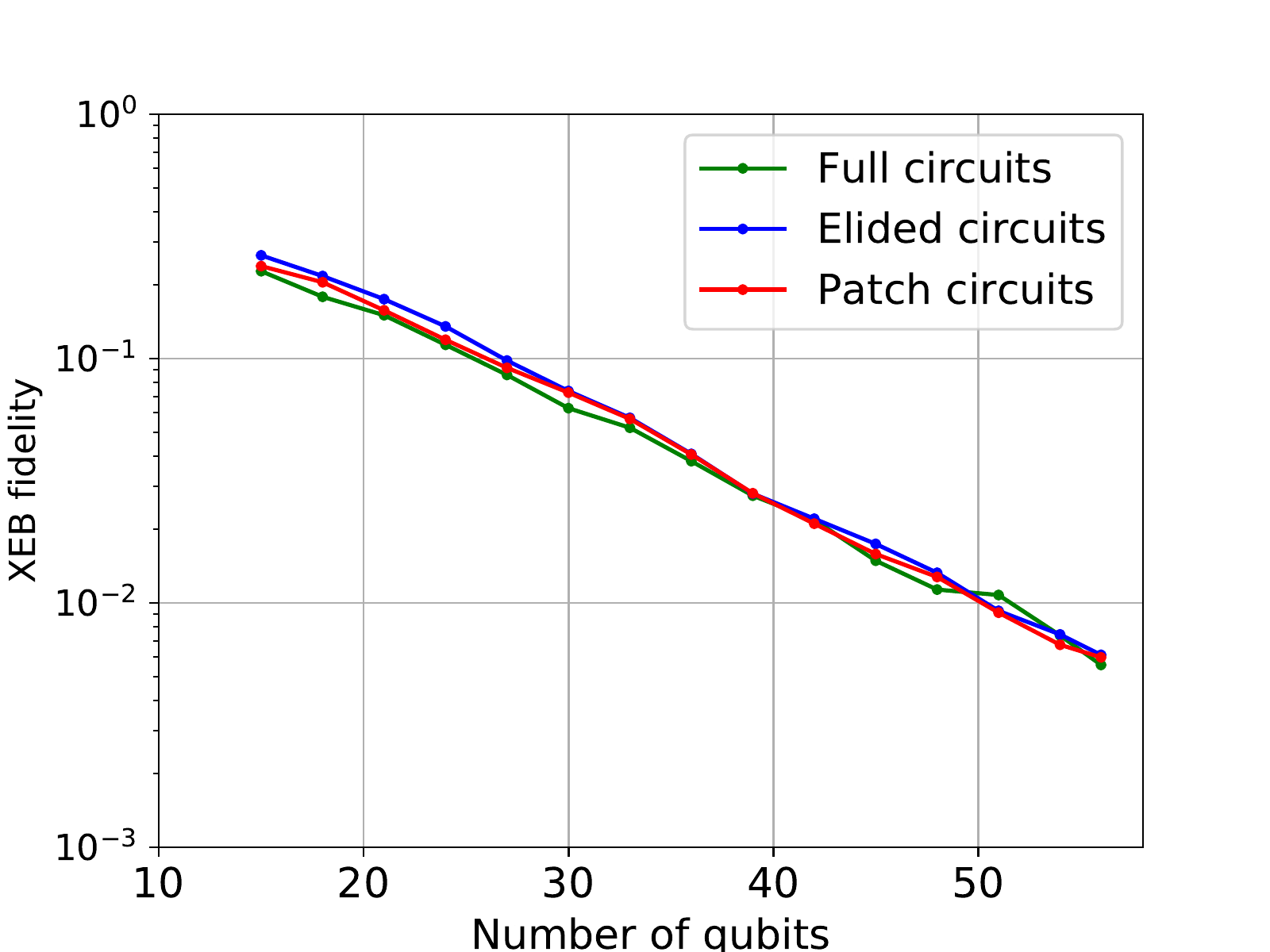}
\end{center}
\caption{\textbf{Performance of patch circuits and elided circuits.} Patch, elided and full circuit XEB fidelity from 15 to 56 qubits with 10 cycles, showing patch and elided circuits yields a fidelity value that is in good agreement with the one obtained with the corresponding full circuits. Each data point is averaged over 6 quantum circuit instances.
\label{XEB_Patch}}
\end{figure}

We evaluate the efficacy of using patch circuits and elided circuits for performance estimation via a direct comparison with full circuits. In Fig.~\ref{XEB_Patch}, the XEB fidelities measured by full, patch and elided circuit of systems from 15 to 56 qubits with 10 cycles are displayed. The fidelities derived from patch and elided circuits exhibit a consistent exponential decay with system size, and are in good agreement with the fidelities obtained with the corresponding full circuits. The average ratio of patch circuit and elided circuit fidelity to full circuit fidelity over all verification circuits are found to be 1.05 and 1.10, with a standard deviation of 8\% and 9\%, dominated by system fluctuations.

\subsection{Distribution of bitstring probabilities}
%After sampling the bitstrings, we calculate the ideal probability of each bitstring. Then we examine the distribution of probabilities.
For a circuit with sufficient depth, the distribution should be consistent with the theoretical prediction. For the bitstrings with fidelity using linear XEB, the theoretical PDF for the bitstring probability $p$ is
\begin{equation}
P_l (x|\hat{F_l} ) = (\hat{F_l} x+(1-\hat{F_l} ))e^{-x}	
\end{equation}
where $x \equiv Dp$ is bitstring probability scaled by the dimension $D$.

The PDF for logarithmic XEB is
\begin{equation}
P_c (x|\hat{F_c}) = (1+\hat{F_c}(e^x-1))e^{x-e^x}	
\end{equation}
where $x \equiv log(Dp)$.

\begin{figure}[!htbp]
\begin{center}
\includegraphics[width=0.9\linewidth]{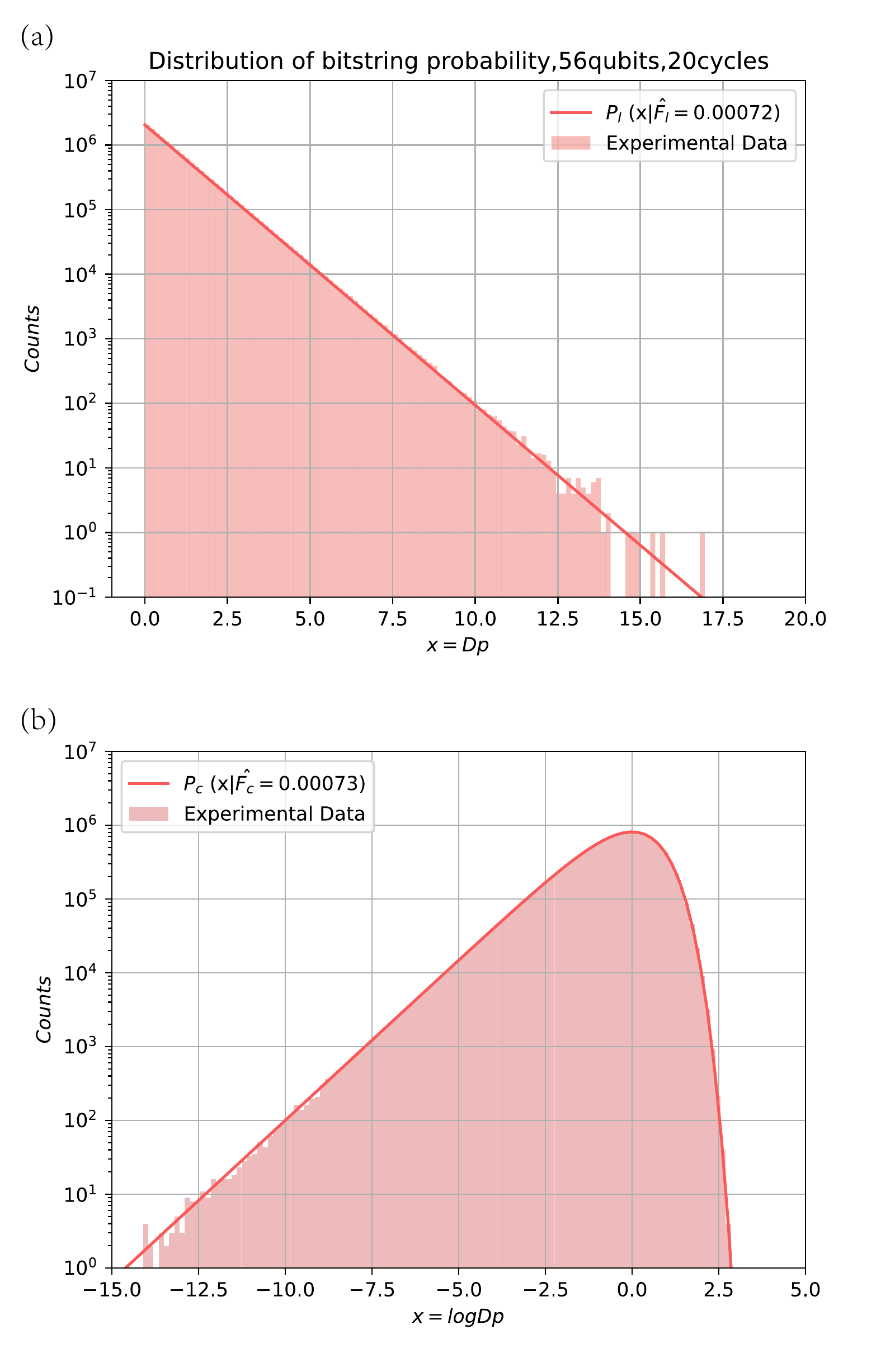}
\end{center}
\caption{\textbf{Distribution of bitstring probabilities from a 56-qubit 20 -cycle circuit.} The theoretical curve is computed with the experimental XEB fidelity and experimental data is counted by histogram.  (a) The theoretical curve $P_l (x|\hat{F_l} )$ and experimental distribution of $Dp$. (b) The theoretical curve $P_c(x|\hat{F_c} )$ and experimental distribution of log($Dp$).}
\label{Fig.1}
\end{figure}

For our elided quantum circuit with 56 qubits and 20 cycles, we sample around $1.9 \times 10^7$ bitstrings from each of 10 instances and calculate the linear XEB $\hat{F_l}$ or logarithmic XEB $\hat{F_c}$. Then we calculate the ideal probability $p$ of each bitstring to check whether it fits the theoretical curve $P_l (x|\hat{F_l} ) $ and $P_c (x|\hat{F_c}) $. The result of one circuit instance is shown in Fig.~\ref{Fig.1}.

\begin{comment}
\begin{figure}[!htbp]
\begin{center}
\includegraphics[width=0.9\linewidth]{fig2}
\end{center}
\caption{\textbf{The Kolmogorov-Smirnov test results.} $D_{KS}$ and p-values are shown for linear XEB (upper) and log XEB (lower).}
\label{Fig.2}
\end{figure}
\end{comment}

We further use Kolmogorov-Smirnov test (K-S test)~\cite{lehmann2006testing} to quantify the agreements between experiment data and theoretical curve. K-S test quantifies a distance between the empirical cumulative distribution and the hypothetic cumulative distribution function, denoted as $D_{KS}$. $D_{KS}$ can be converted to $p$-value according to  Kolmogorov distribution function. We use $p$-value to judge whether the hypothesis is good. We use kstest function in the scipy package~\cite{jones2001scipy} with the null hypothesis of $F=\hat{F}$. For comparison, we implement KS-test with the null hypothesis of $F=0$ as well.

%The maximum $p$-value of 9 circuit instances is \hhl{$F=0$ is 0.045}. We can confidently reject the null hypothesis \hhl{$F=0$ with more than $95\%$} confidence. For $F=\hat{F}$, the $p$-value is significantly greater than zero. We can conclude that the experiment data shows good agreements with theoretical prediction.
We measure the XEB and do KS-test with bitstrings from ten instances combined. The linear XEB and logarithmic XEB  fidelities from the combined bitstrings are $\hat{F_l}=6.60 \times 10^{-4}$ and $\hat{F_c}=5.80 \times 10^{-4}$, respectively. The KS-test results are shown in table~\ref{tab1}. With null hypothesis of $F=0$, the p-value is around $9.7\times 10^{-11}$. We can reject the null hypothesis confidently.

\begin{table}[!htbp]
\begin{tabular}{|c|c|c|c|c|l|l|}
\hline
\multicolumn{1}{|l|}{} & \multicolumn{2}{c|}{linear XEB}     & \multicolumn{4}{c|}{log XEB}                               \\ \hline
hypothesis             & $F=\hat{F_l}$ & $F=0$               & $F=\hat{F_c}$   & \multicolumn{3}{c|}{$F=0$}               \\ \hline
$p$-value                & 0.23          & $9.7\times 10^{-11}$ & 0.66            & \multicolumn{3}{c|}{$9.7\times 10^{-11}$} \\ \hline
\end{tabular}
\caption{\textbf{Results of combined Kolmogorov-Smirnov test for random circuits with 56-qubit, 20 cycles.}}
\label{tab1}
\end{table}

%\subsection{XEB uncertainties}
\subsection{Statistical uncertainties}
%The uncertainties of XEB measurements include statistical uncertainties and systematic uncertainties.
We estimate the statistical uncertainty of XEB measurements with standard error-on-mean formula
\begin{equation}\begin{array}{l}
\hat{\sigma}_{F_{l}}=D \sqrt{\operatorname{Var}(p) / N_{s}}, \\
\hat{\sigma}_{F_{c}}=\sqrt{\operatorname{Var}(\log Dp) / N_{s}},
\end{array}
\label{eq.3}
\end{equation}
where Var($x$) is the variance estimator of sample $\{x_i\}$. We use inverse-variance weighting to estimate the fidelity and statistical uncertainty of all nine 56-qubit and 20-cycle random circuits, yields the results of $\hat{F_{l}} = (6.62\pm 0.72)\times 10^{-4}$ for linear XEB and $\hat{F_{c}} = (5.82\pm 0.92)\times 10^{-4}$ for logarithmic XEB. The theoretical prediction of the statistical uncertainty of linear XEB $\hat{F_{l}}$ and logarithmic XEB  $\hat{F_{c}}$ are

\begin{equation}\begin{array}{l}
\hat{\sigma}_{F_{l}}=\sqrt{\left(1+2 F-F^{2}\right) / N_{s}},\\
\hat{\sigma}_{F_{c}}=\sqrt{\left(\pi^{2} / 6-F^{2}\right) / N_{s}},
\end{array}\end{equation}
which are calculated as $\hat{\sigma}_{F_{l}} = 7.2\times 10^{-5}$ and $\hat{\sigma}_{F_{c}} = 9.2\times 10^{-5}$, respectively, indicating good agreements between experiment and theory.

\begin{figure}[tbp]
\begin{center}
\includegraphics[width=0.9\linewidth]{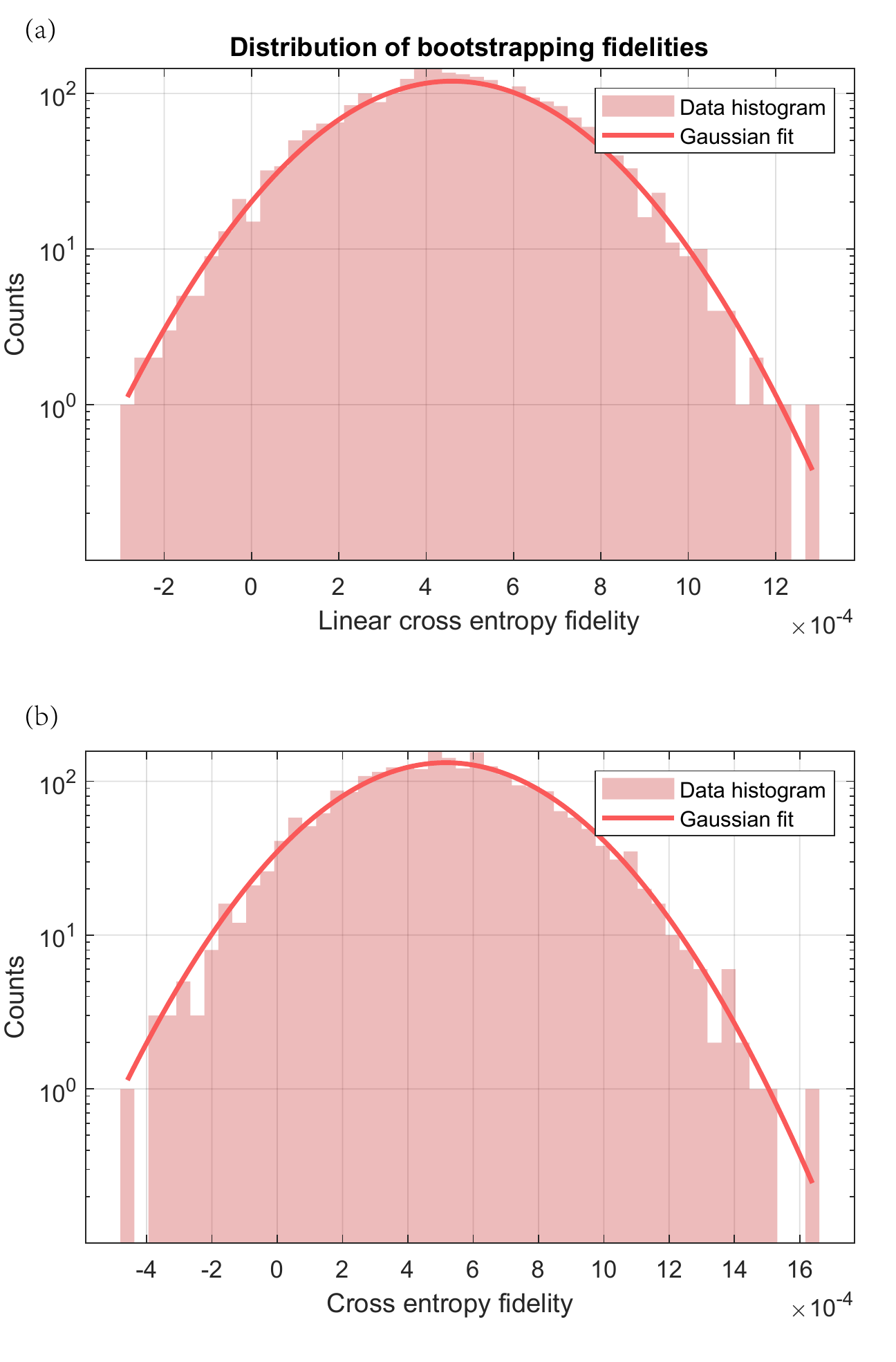}
\end{center}
\caption{\textbf{Fidelity distribution of bootstrap samples.} (a) Linear XEB distribution. (b) Log XEB distribution.}
\label{Fig.3}
\end{figure}

We further use bootstrap method to verify the estimate of statistical uncertainties. For a 56-qubit 20-cycle circuit, we sample  bitstrings from experiment and calculate their ideal probabilities to be the original sample. We acquire 2500 bootstrap samples from the original one and compute $F=\hat{F_l}$, $F=\hat{F_c}$ of every sample. We expect the set of $F=\hat{F}$ follow Gaussian distribution under the central limit theorem. The fidelity distribution of bootstrap samples with Gaussian fit are shown in Fig.~\ref{Fig.3}.
We compare the statistical uncertainty obtained from equation (\ref{eq.3}), Gaussian fit and the standard deviation of the bootstrap distribution. The results are $2.41, 2.40, 2.36(\times 10^{-4})$ for $\hat{\sigma}_{F_{l}}$ and $3.09, 3.16, 3.07(\times 10^{-4})$ for $\hat{\sigma}_{F_{c}}$. It indicates great consistency within $3\%$ relative difference.

\section{classical simulation}
\subsection{The efficiency of Schr{\"o}dinger simulator}
In the hybrid Schr{\"o}dingerFeynman algorithm (SFA) simulator, each path is simulated using the Schr{\"o}dinger algorithm (SA). The Schr{\"o}dinger algorithm is a full state vector simulator for simulating quantum circuit. It computes all $2^n$ amplitudes, where $n$ is the number of qubits. In this section, we will test the performance of the Schr{\"o}dinger simulator.

Following the work in Ref.~\cite{arute2019quantum}, gate fusion~\cite{smelyanskiy2016qhipster} and single precision arithmetic are used in our simulator. We simulate quantum circuits on a single node server that has 1536 GB memory and four CPUs (Intel-Xeon-Gold-6254, 3.1G) with 18 cores each. To compare with the performance of the Schr{\"o}dinger simulator in Ref.~\cite{arute2019quantum}, we test random circuits with different sizes at depth 14. The run times are listed in Table ~\ref{tab:Schrodinger simulator}. Results show that the performance of our simulator is basically the same as that of Google~\cite{arute2019quantum}.

\begin{table}[!htbp]
\centering
\newcommand{\tabincell}[2]{\begin{tabular}{@{}#1@{}}#2\end{tabular}}
\begin{tabular}{|c|c|c|}
\hline
number of qubits   & \tabincell{c}{run time in seconds\\(ours)} & \tabincell{c}{run time in seconds\\(Google's~\cite{arute2019quantum})}\\
\hline
$30 $ & 24   &  NA   \\
$32$ & 93   &   111  \\
$33$ & 190  &   NA  \\
$34$ & 362  &   473  \\
$36$ & 1836  &  1954 \\
\hline
\end{tabular}
\caption{\textbf{Circuit simulation run times using Schr{\"o}dinger simulator.}  \label{tab:Schrodinger simulator} }
\end{table}

\subsection{Tensor network simulator}

Tensor network contraction (TNC) algorithm translates the task of computing a single or a branch of amplitudes into contracting a tensor network, where each tensor corresponds to a quantum gate operation. The complexity of TNC algorithm is controled by the largest intermediate tensor which appears during the contraction, which is also related to the tree width of the line graph corresponding to the tensor network. As a result the ultimate performance of TNC algorithm is determined by the underlying tensor contract path. In this work we use the python package cotengra to find an optimal tensor contraction path, which has been demonstrated to be able to reproduce the state of the art results in Ref.~\cite{HuangChen2020,pan2021simulating}. For each random quantum circuit, we repeated the 100 path search procedure to determine the optimal tensor contraction path and use its corresponding number of floating point operations for contraction as the estimated computational cost.

In the main text, the modified frugal rejection sampling~\cite{arute2019quantum, markov2018quantum} with acceptance probability $1$ is used to estimate the cost of tensor network algorithm on specific circuits~\cite{VillalongaMandra2019,HuangChen2020}. In Ref.~\cite{pan2021simulating}, a subspace sampling trick is proposed, which has been proven to be able to fool linear XEB. However, it is easy to distinguish this type of sampling, since some positions in the sampled bitstrings are fixed to 0, such as $x_1x_2x_3x_400000000$, where $x_i$ is a variable bit.

In Ref.~\cite{HuangChen2020,pan2021simulating}, the actual cost of tensor network contraction using GPU is provided, which could be regard as reference to estimate the actual cost of our circuit in supercomputer, such as Summit (Summit has 27,648 GPUs in total, but Fugaku has no GPUs, therefore we utilize Summit to determine the actual cost). Take the results in Ref.~\cite{HuangChen2020} as an example, it would cost 833.75s to generate one perfect sample for a tensor network with $6.66\times 10^{18}$ contraction cost using Summit. Thus, it would cost $\frac{{1.10 \times {{10}^{22}}}}{{6.66 \times {{10}^{18}}}} \times 833.75\text{s}{\rm{ = }}15.9$ days and $\frac{{2.08 \times {{10}^{24}}}}{{6.66 \times {{10}^{18}}}} \times 833.75\text{s}{\rm{ = }}8.24$ years to reproduce the same results as the 53-qubit 20-cycle circuit in Ref.~\cite{arute2019quantum} and our 56-qubit 20-cycle circuit using Summit.

\subsection{Computational cost estimation of SFA for the sampling task}
Here, we estimate the computational cost of simulating 56 qubit full-circuits using SFA. For the 56-qubit random circuit with 20 cycles, there are 42 gates cross the cut. The iSWAP-like gate has a Schmidt rank of 4. However, the first one and last three iSWAP-like gates can be simplified to cphase with a Schmidt rank of 2. In the case of simulating quantum circuit with $100\%$ fidelity, a total of $ 4 ^ {38} \times 2^4$ paths must be calculated. By using the technique of \textit{prefix}~\cite{markov2018quantum} to optimize the simulator, we set a prefix of 35 cross gates (a cross gate can be simplified to a cphase gate), thus requiring $4^{34}\times 2^{1}$ separate runs.

We simulate the quantum circuit for the first 10 prefix values on a single node server that has 1536 GB memory and four CPUs (Intel-Xeon-Gold-6254, 3.1G) with 18 cores each.
%The execution time for these \hhl{100 prefix} are shown in \hhl{Fig.~\ref{Qsimh_execution_time}}, and
The average execution time for each prefix is 19560 seconds using single core and single thread. Thus, it is estimated that it will consume $1.06\times 10^{18}$ core (two hyperthreads) hours to simulate the 56-qubit 20-cycle circuit with 0.0662\% fidelity. Table~\ref{tab:qsimhruntime} also shows the extrapolated run times for Google's 53-qubit 20-cycle circuit ~\cite{arute2019quantum} using our SFA simulator ($8.9\times 10^{13}$ core hours).

%\begin{figure}[htbp]
%\begin{center}
%\includegraphics[width=1.0\linewidth]{Qsimh_execution_time}
%\end{center}
%\caption{\textbf{Execution time for a 60 qubit circuit with 20 cycles.}  We simulate the circuit for the first %10 prefix values. The average execution time is calculated to be \hhl{XXX} seconds.
%\label{Qsimh_execution_time}}
%\end{figure}

\begin{table}[htbp]
\centering
\newcommand{\tabincell}[2]{\begin{tabular}{@{}#1@{}}#2\end{tabular}}
\begin{tabular}{|c|c|c|c|c|}
\hline
\# of qubits & cycle & number of paths & fidelity & run time (years) \\
\hline
$53 $ & 20   &  $4^{31}\times2^4$ & 0.224$\%$ &  1,332  \\
$56 $ & 20   &  $4^{38}\times2^4$ (balance) & 0.0662$\%$ &  15,887,738   \\
$56 $ & 20   &  $4^{35}\times2^6$ (imbalance) & 0.0662$\%$ &  8,612,623  \\
\hline
\end{tabular}
\caption{\textbf{Run times of SFA using 7,630,848 CPU cores} (the most powerful supercomputer Fugaku has a total of 7,630,848 cores).  \label{tab:qsimhruntime} }
\end{table}

\begin{figure}[tbp]
\begin{center}
\includegraphics[width=0.85\linewidth]{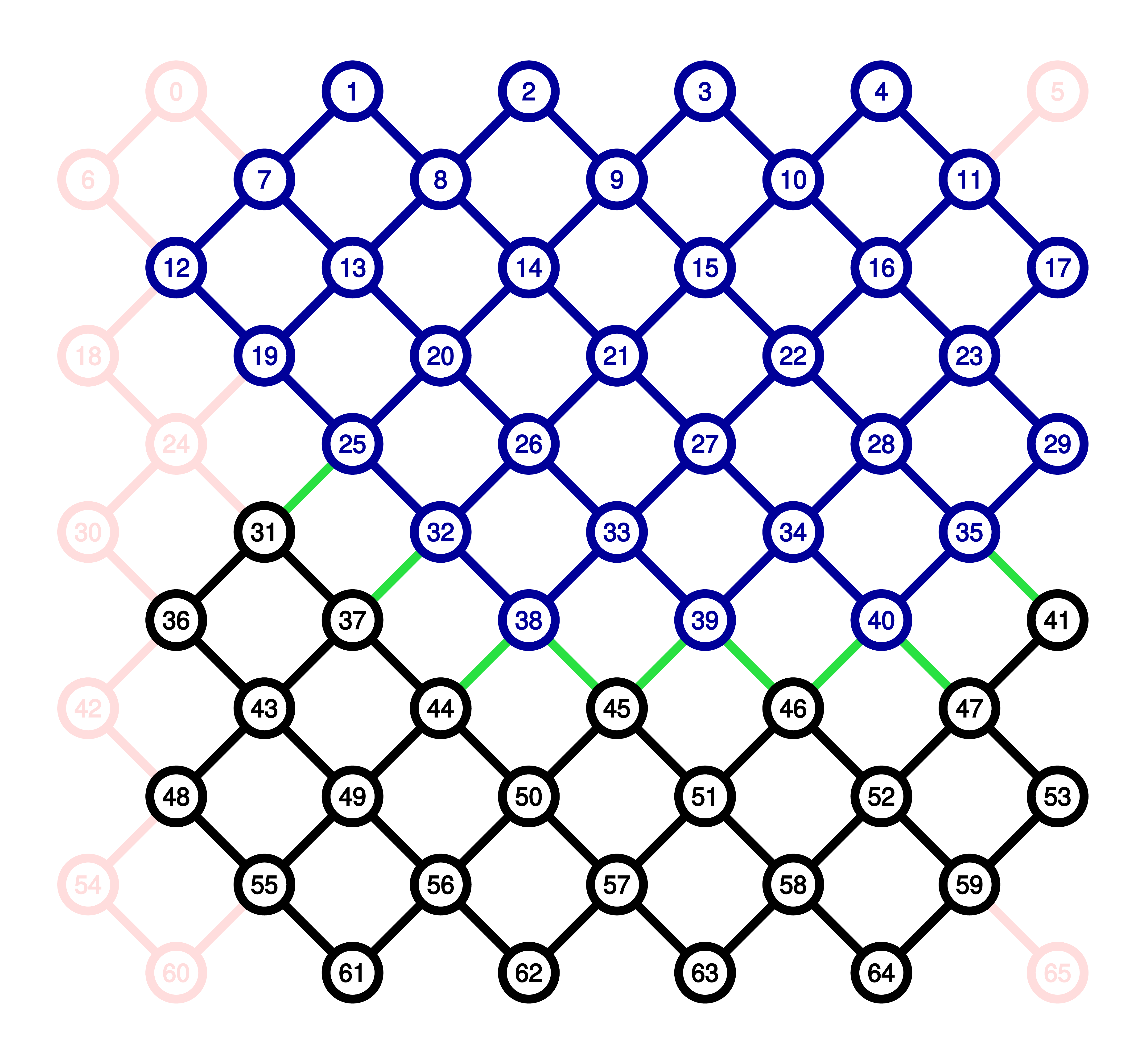}
\end{center}
\caption{\textbf{Imbalance cut.} We set the allowed imbalance of two partitions to 20 ($n_1-n_2\le20$), and employ the search strategy in Section IV to find the optimal cut. The number of qubits in the two partitions are 32 and 24, respectively.
\label{imbalanced_cut}}
\end{figure}

We note that in the results in Table~\ref{tab:qsimhruntime}, we also provide a result of using the imbalance cut shown in Fig.~\ref{imbalanced_cut}, which consumes less time ($5.76\times10^{17}$ core hours) but consumes more storage space. In addition, we did not consider the DCD formation in our estimation. The DCD formation appears twice in our 56-qubit 20-cycle circuit, and also twice in Google's 53-qubit 20-cycle circuits. Thus, after considering the simulating speedup by DCD formation, we have
\begin{align}
\frac{{{T_{56\_20}}}}{{{T_{53\_20}}}} = \frac{{\frac{{8612623~\text{years}}}{{{4^2}}}}}{{\frac{{1332~\text{years}}}{{{4^2}}}}} \approx 6466
\end{align}
where ${{T_{56\_20}}}$ and ${{T_{53\_20}}}$ are the run time of simulating our 56-qubit 20 cycle circuit and Google's 53-qubit 20-cycle circuit, respectively. That is, using the SFA algorithm, the computational cost of simulating our 56-qubit 20-cycle circuit is about 6466 times that of Google's 53-qubit 20-cycle circuit.

\subsection{Classical speedup for imbalanced gates}
The iSWAP-like$(\theta ,\phi )$ gate has the following Schmidt singular values:
\begin{align}
{\lambda _1} &= \sqrt {1 + 2 \cdot |cos(\phi /2)cos\theta | + co{s^2}\theta },\\
{\lambda _2} &= \sin (\theta ),\\
{\lambda _3} &= \sin (\theta ),\\
{\lambda _4} &= \sqrt {1 - 2 \cdot |cos(\phi /2)cos\theta | + co{s^2}\theta },
\end{align}

We have $\theta  \approx \pi /2$ and $\phi  \approx \pi /6$ in our experiment, so ${\lambda _i} \approx 1,{\kern 1pt} {\kern 1pt} \forall i \in \{ 1,2,3,4\} $. When simulating the random circuit sampling with a target fidelity using SFA simulator, inbalanced iSWAP-like gates can provide acceleration for the simulation.  According to the Ref.~\cite{arute2019quantum},~
%, assume that all $g$ gates have the same values of $\theta$ and $\phi$.
if $100\%$ fidelity is required, a total of $ 4 ^ g $ paths must be calculated. However, given a target fidelity, $F$, one need consider only the top $S$ paths with the highest weight, making
\begin{align}
F = \sum\nolimits_{i = 1}^S {\frac{{{W_i}}}{{{4^g}}}},
\end{align}
where ${W_i} = \lambda _{{i_1}}^2\lambda _{{i_2}}^2 \ldots \lambda _{{i_g}}^2$ is the weight of each path arising from this decomposition. For comparison, if all iSWAP-like gates are balanced with $\theta {\rm{ = }}\pi /2$, the number of paths needs to be considered is $F \times {4^g}$. Thus, imbalanced gates provides a
speedup equal to $\frac{S}{{F \times {4^g}}}$.

\begin{figure}[htbp]
\begin{center}
\includegraphics[width=\linewidth]{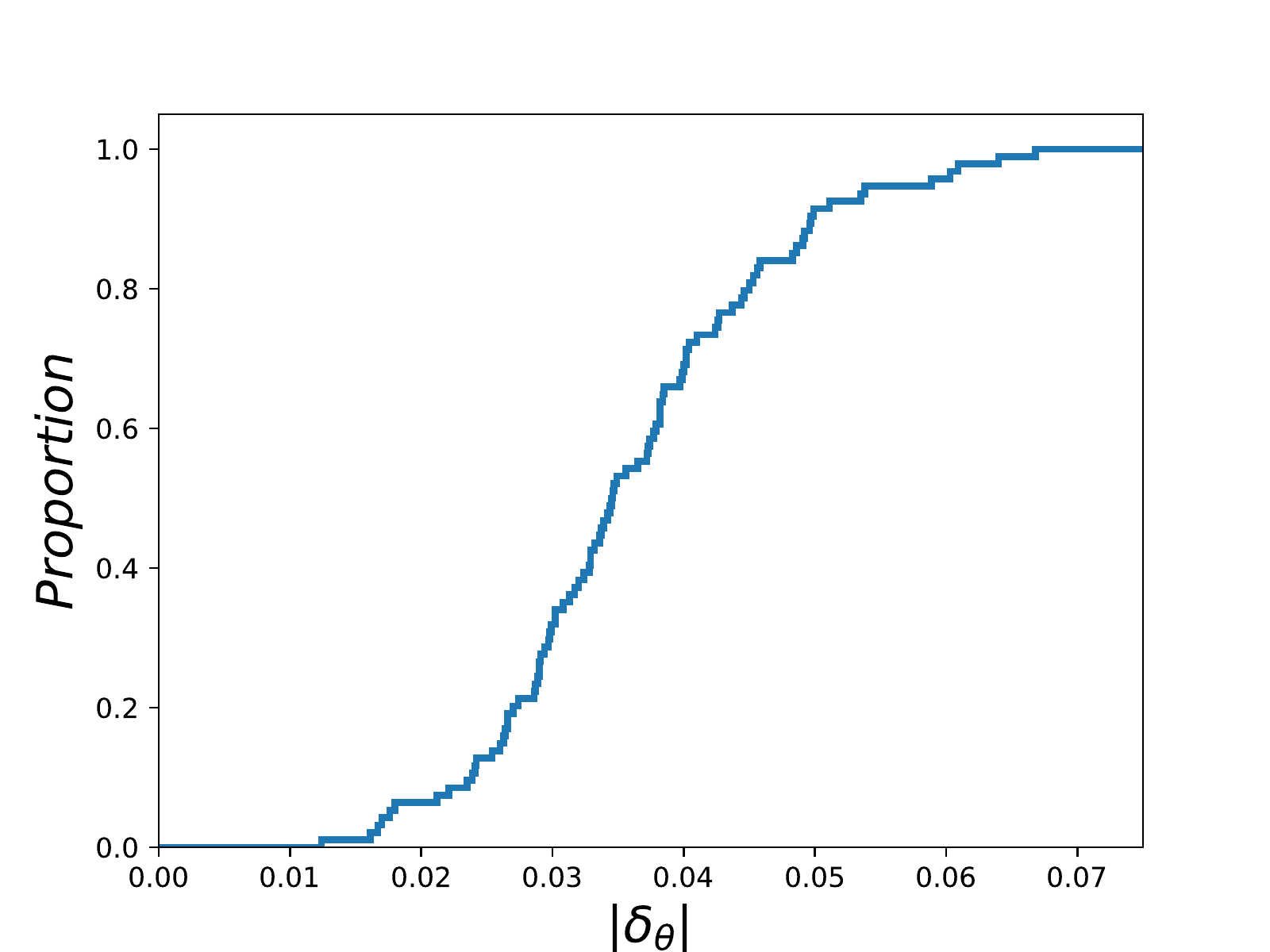}
\end{center}
\caption{\textbf{Cumulative probability distribution of $|{\delta _\theta }|$ for iSWAP-like gates.} In experiment, we have an average $|\delta_{\theta}| \approx 0.036$ for all iSWAP-like gates. The resulted speedup of classical simulation is less than an order.
\label{fSimgatesdistribution}}
\end{figure}

\begin{figure}[!htbp]
\begin{center}
\includegraphics[width=1.1\linewidth]{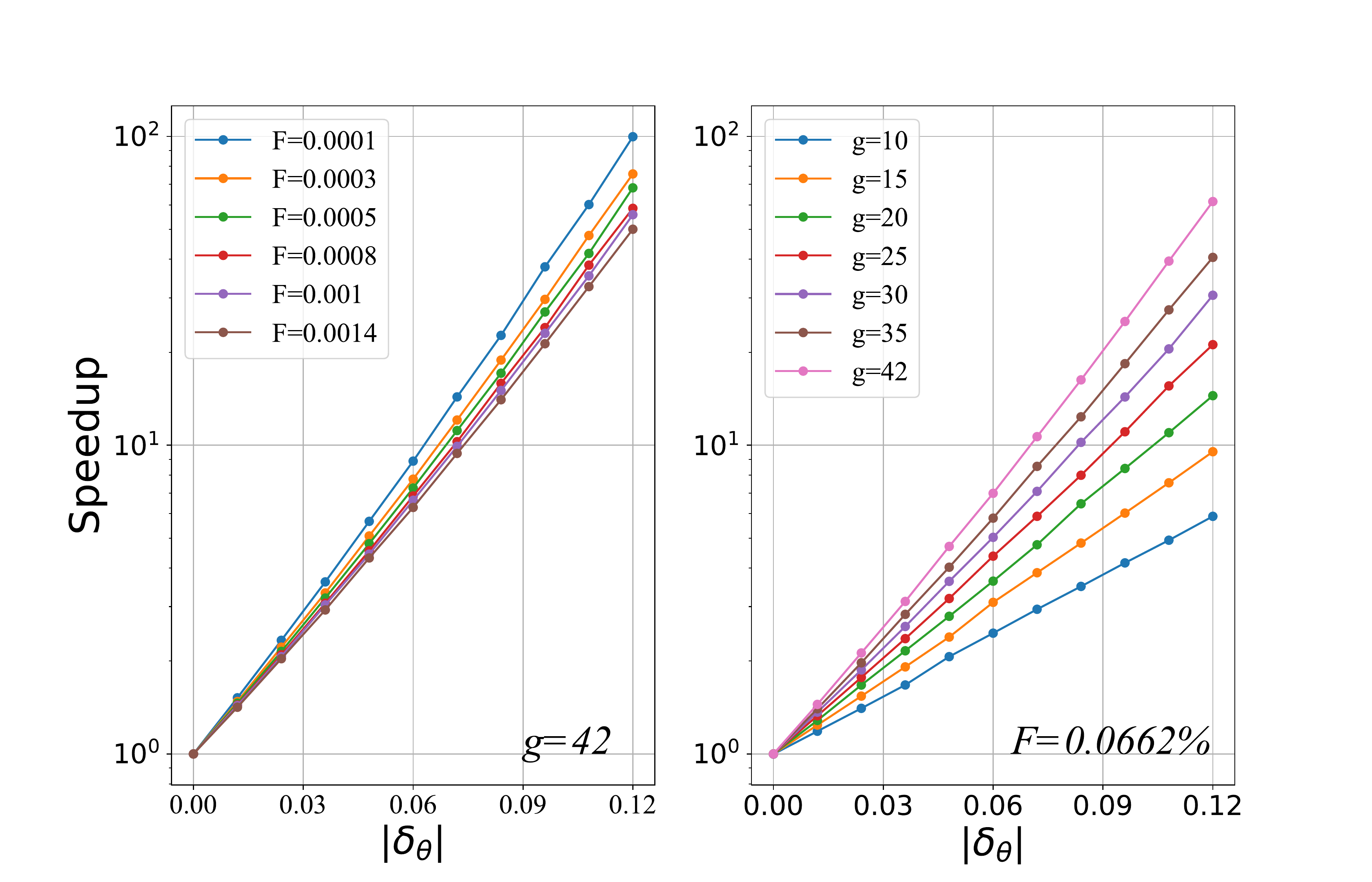}
\end{center}
\caption{\textbf{Classical speedup given by the imbalance gates.} We assume all iSWAP-like gates deviate from $\pi/2$ by the same $\delta_{\theta}$ and calculate the speedup with given $g$ and $F$. \textit{Left:} speedup with varied fidelity $F$ and fixed $g=42$. \textit{Right:} speedup with varied g and fixed $F=0.0662 \%$.
\label{imbalancedgates}}
\end{figure}

\begin{figure*}[!htbp]
\begin{center}
\includegraphics[width=1\linewidth]{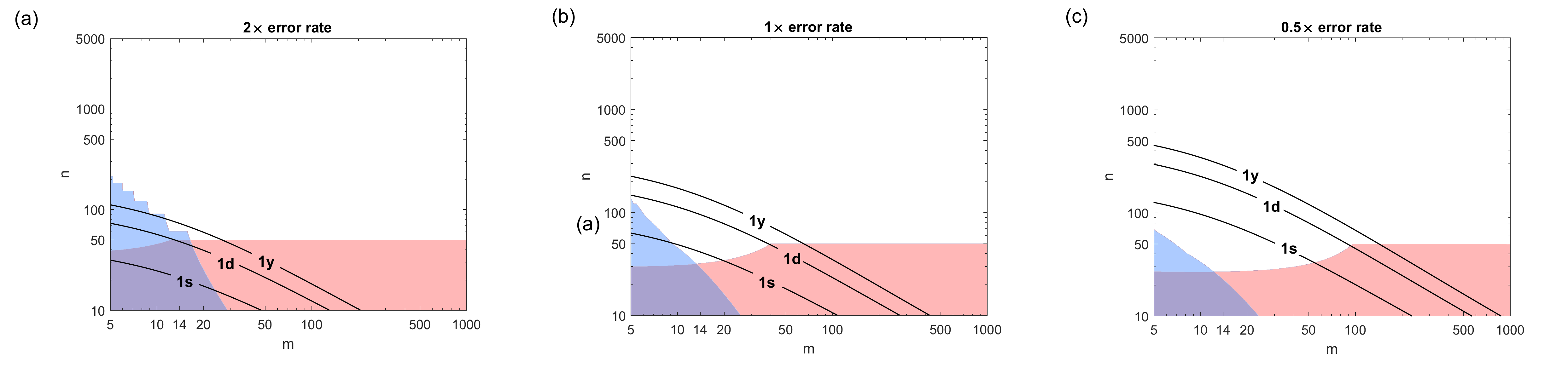}
\end{center}
\caption{\textbf{Runtime advantage region of different error rates.} The colored regions indicate classical runtime advantage beyond the quantum computer within the limit of supercomputer memory. Runtime advantage depends on the circuit depth $m$ and number of qubits $n$. Red and blue region indicates SA advantage and SFA advantage, respectively. The black contours indicate runtime of quantum computer.
%Our experiment circuits are marked with red pentacles.
From (a) to (c), the average error rates of quantum computer are $2\times, 1\times, 0.5\times$ our current experimental error rates.}
\label{Fig.6}
\end{figure*}

In our experiment, the iSWAP-like gate that need to be decomposed is iSWAP-like$(\pi/2 \pm {\delta _\theta },\phi  \approx \pi /6)$, and $|{\delta _\theta }|$ has values of around 0.036 radians (see \hhl{Fig.~\ref{fSimgatesdistribution}}). The speedup can be achieved from the imbalanced gates is shown in \hhl{Fig.~\ref{imbalancedgates}}. For the case of random circuit sampling with $n=56$ qubits and $m=20$ cycles, the fidelity is around $F=0.0662\%$ and the number of iSWAP-like gate need to be decomposed is $g=42$ using SFA simulator, the speedup estimated would be well below an order of magnitude (see \hhl{Fig.~\ref{imbalancedgates})}.

\subsection{Quantum runtime advantage region}

Tensor network algorithms may have higher simulation efficiency at low depths. However, for large-scale quantum circuit with high depth, SFA is still the most efficient algorithm at present. In this section, we first analyze the scaling of the computational cost of SA and SFA, then we give a rough estimate of the quantum runtime advantage region. There may be other classical simulator with better performance, but this estimate is just to illustrate the importance of improving the fidelity of quantum operations, including quantum gates and readouts.

%To simulate the quantum circuit, we have used SA and SFA simulators. SA simulator computes all state vectors and gates with matrix manipulation. In SFA, we cut the system into several patches and use SA to compute every patch. On the cut, we decompose the gates as different paths and add them up like Feynman integral. \\

%We follow the procedure in Ref.~\cite{arute2019quantum}~\cite{zlokapa2020boundaries} to estimate the cost of SA and SFA.
For a $n$-qubit quantum circuit with $m$ cycles, the runtime of SA is estimated as
\begin{equation}
T_{\text{SA}} = C_{\text{SA}} ^{-1} \cdot mn \cdot 2^n
\end{equation}
where the constant $C_{\text{SA}}$ is fit to the actual runtime of a state-of-the-art supercomputer. SA needs $2^{n+1}$ bytes to store the complex state vector. Considering that state-of-the-art supercomputers have less than 3 PB of memory, the maximum number of qubits that can be simulated using the SA simulator is 51 at most.

For SFA, the runtime is proportional to the number of paths and the time to simulate patches. In Ref.~\cite{arute2019quantum}, the circuit is cut into 2 patches.
%They estimate the number of path needed to simulate as $2^{2Bm\sqrt{n}}$.
However, for larger-scale quantum circuits, we can cut the circuit into more than 2 patches. According to the Ref.~\cite{zlokapa2020boundaries},  we need to simulate the $2^{kpBm\sqrt{n}}F$ paths for $p$ patches, where  $k=1/2+1/p$, $B=0.24$, and $F$ is the fidelity of circuit. The time to simulate each patch scales with $2^{n/p}$. In addition, we compute the partial amplitudes of min$(F^{-2},2^n)$ bitstrings after simulating each patch. In total, the runtime is~\cite{zlokapa2020boundaries}
\begin{equation}
T_{\text{SFA}}=C_{\text{SFA}}^{-1} 2^{kpBm \sqrt{n}} F (p 2^{n/p} + \text{min}(F^{-2},2^n)),
\end{equation}
where $C_{\text{SFA}}$ is fit to the actual runtime of a state-of-the-art supercomputer. We can optimize the runtime with  $F^{-2} = p 2^{n/p}$ for $n>\text{log}_{2}(p) /(1-1/p)$. SFA requires $2p2^{n/p}$ bytes per path. Assuming that each path is simulated by a single core, then we estimate the total memory footprint to be $10^6 \cdot 2p2^{n/p}$ for a supercomputer with 1M cores. Memory usage limits the number of patches, which should be taken into account when optimizing the runtime of SFA.

%As for verifiable circuits, the number of paths needed reduces because of the wedges.
%\begin{equation}
%T_{SFA,v}=C_{SFA,v}^{-1} 2^{\frac{4}{7} kpBm \sqrt{n}} F (p 2^{n/p} + min(F^{-2},2^n)),
%\end{equation}
%where the subscript $v$ represents verifiable circuit.

We compare the runtime of classical simulation with quantum runtime to give a rough estimate of the quantum runtime advantage region. For this purpose, we use the classical fitting constants in Ref.~\cite{arute2019quantum} to continue the discussion.
\begin{equation}
\begin{aligned}
&C_{\text{SA}}= 0.015 \times 10^6 \text{GHz}  \\
&C_{\text{SFA}}= 3.3 \times 10^6 \text{GHz}
\end{aligned}
\label{eq.9}
\end{equation}

The runtime of quantum computer is proportional to the number of samples. To ensure the standard deviation is less than the fidelity, i.e. $\sigma \leq F_{\text{XEB}}$, at least $1/F^2$ samples are required. The runtime of quantum computer scales as
\begin{equation}
T_{\text{Q}} = \frac{1}{C_{\text{QC}} \cdot F^2}
\end{equation}
where $C_{\text{QC}}=\frac{1}{230}$MHz is the actual sampling rate of our quantum computer.

To compare the runtime of different methods, we fit constants in Eq.~\ref{eq.9} and $F$ from Eq.~\ref{fidelitycal} into the runtime estimation discussed above and optimize the runtime of SFA. In addition, the memory constraint is taken into account. The result is shown in Fig.\ref{Fig.6}. The quantum advantage region (white) indicates the ability of quantum computer beyond classical computer. It is worth mentioning that the quantum advantage region enlarges rapidly when error rates decline, indicating the importance of low error rates.
%But the contours constrain the real quantum advantage region due to the long quantum runtime.

\bibliographystyle{apsrev4-1}
\bibliography{references}